\documentclass[12pt,onecolumn]{article}

\usepackage{latexsym}
\usepackage{setspace}
\usepackage{a4wide}
\usepackage{amsmath}
\usepackage{amsfonts}
\usepackage{amscd}
\usepackage{amssymb}
\usepackage{amsbsy}
\usepackage{mathrsfs}
\usepackage{yfonts}
\usepackage{bbm}
\usepackage[dvips]{epsfig,graphics}
\usepackage{epsfig}
\usepackage{graphicx}
\usepackage{psfrag}
\usepackage{upgreek}
\usepackage{undertilde}
\usepackage{color}
\usepackage{leftidx}

\DeclareMathAlphabet{\mathpzc}{OT1}{pzc}{m}{it}

\begin{document}

\title{Stress intrepretation of graphene's $E_{2g}$ and $A_{1g}$ vibrational modes: theoretical analysis}

\author{D. Sfyris, G.I. Sfyris, C. Galiotis}

\maketitle

\begin{abstract}

Laser Raman spectroscopy is a powerful non-destructive technique for monitoring the local stress variations within specimens of very small size and nowadays it is an integral part for graphene research. Converting spectroscopic data collected during a Raman experiment into macroscopic quantities requires good understanding of the connection between mechanical fields and the microscopic dynamics. We here focus on only one graphene ring and examine to which stress tensor components the $E_{2g}$ and the $A_{1g}$ vibration mode of graphene correspond. These modes are typically related with the G-peak and the D-peak, respectively, and are strongly related to the stress distribution along the specimen. We adopt the theoretical framework of Admal and Tadmor (\cite{Admal-Tadmor2010}) for the macroscopic definition of the Cauchy stress tensor and we introduce into this framework the $E_{2g}$ and the $A_{1g}$ as appropriate perturbations. We use these perturbations to the stress tensor expression and evaluate which stress tensor components are related to each vibrational mode. This approach, though qualitative in nature, incorporates all the main physics and reveals that $E_{2g}$ and $A_{1g}$ vibration modes should be related to shear as well as axial stress components when graphene is at rest (i.e. no external applied loading). To bring our framework closer to more concrete results, we evaluate the instantaneous Hardy stress tensor for a pair potential which correspond to the $E_{2g}$ and $A_{1g}$ modes at rest. Our analysis expands to take into account an applied external tensile field. Taking the armchair direction to be along the x-axis, when tension applies along the armchair direction, it is the axial $\sigma_{11}$ stress component which dominates over $\sigma_{12}, \sigma_{22}$, which are of smaller order. When tension is along the zig-zag direction, it is the axial $\sigma_{22}$ stress component that dominates over $\sigma_{12}, \sigma_{11}$. When tension is at an arbitrary direction between the armchair and the zig-zag direction, all stress components are of the same order and should all be taken into account even at small strains. Specifically, as the inclination of the applied tension field approaches 45$^0$, the shear component increases and at value of angle 45$^0$ it achieves its maximum value, being of the same order with its axial counterparts.   

\end{abstract}


\section{Introduction}

Laser Raman spectroscopy is a powerful non destructive technique for monitoring the local stress variations within specimens of very small size. The Raman microprobe used in these experiments can provide a 500nm spatial resolution \footnote{Raman spectroscopy as an optical technique cannot be less that the wavelength of light in the visible range. Here we do not cover special techniques such as TERS that can go down in scale.} thus giving information from the submicron scale. In recent years Raman spectroscopy became an integral part of graphene research: it is used to determine the number and orientation of layers, the quality and types of edges, the effects of perturbations, disorder, doping as well as thermal properties (\cite{Ferrari-Basko2013,Ferrari2007,Saitoetal2011,Ferrarietal2006,Ferrari-Robertson2001}). It can also be used to measure the effect of strain (\cite{Mohhiuddinetal2009,Franketal2010,Tsouklerietal2009,Franketal2011,Androulidakisetal2015,Zdetsisetal2009,Zervakietal2014,Siafakaetal2015}) as well as to infer failure mechanisms in embedded graphene (\cite{Androulidakisetal2014}).

Essentially, Raman spectroscopy emerged as the main technique to probe graphene's phonons (\cite{Sahataetal1988,Ferrari2007,Ferrarietal2006,Ferrari-Robertson2001,Saitoetal2011}). Strain modifies the crystal phonons with tensile strain usually resulting in mode softening and the opposite for compressive strain. The rate of these changes is summarized in the Gruneisen parameter which also determines the thermomechanical properties (\cite{Mohhiuddinetal2009}). The G-peak corresponds to the doubly degenerate $E_{2g}$ phonon at the Brillouin zone center. The D-peak is due to the breathing modes of $sp^2$ rings and requires a defect for its activation: it comes from transverse optical phonons around the K point of the Brillouin zone. There is a strong relevance concerning these peaks and mechanical stresses for graphene (\cite{Androulidakisetal2015,Galiotisetal2015}). Earlier approaches on liquid crystal polymer fibers and the convertion of spectroscopic data into stress-strain curves can be seen in the work of Vlattas and Galiotis (\cite{Vlattas-Galiotis1994}).

Converting spectroscopic data into macroscopic quantities requires good understanding of the connection between mechanical fields and the microscopic dynamics. Transfer of information between the discrete model, through the scales to the continuum model, is not a a trivial task since quantities stochastic in nature appear naturally into play. From the continuum point of view, there is a very rich literature regarding the construction of continuum fields for an atomistic system in order to ensure a smooth transfer of information through the scales. Admal and Tadmor (\cite{Admal-Tadmor2010,Admal-Tadmor2011}) in a recent work present a unified framework for the definition of the Cauchy stress tensor based on the Irving and Kirkwood (\cite{Irving-Kirkwood1950}) and Noll (\cite{Noll1955,Lehoucq-Lilienfeld2010}) procedure (hereafter named as IKN procedure) which leads to all of the major stress definitions (virial stress, Tsai traction \cite{Tsai1979} and the Hardy stress \cite{Hardy1982,Murdoch1982,Murdoch2003,Murdoch2007,Murdoch-Bedeaux1993,Murdoch-Bedeaux1994}). Their approach (\cite{Admal-Tadmor2010,Admal-Tadmor2011}) has two steps: firstly they give a pointwise stress tensor using the IKN procedure which contain a statistical mechanics average over the canonical ensemble. Essentially, they generalize the earlier approaches of IKN originally limited to pair potentials to arbitrary multibody potentials. Secondly, they average this pointwise stress field over some spatial region to obtain a true macroscopic quantity. 

Their outcome (\cite{Admal-Tadmor2010,Admal-Tadmor2011}) is a stress tensor $\boldsymbol \sigma$ which is non-unique and symmetric and has two parts: a dynamic part and a kinetic part. The kinetic part reflects the momentum flux associated with the vibrational kinetic energy portion of the internal energy. The source of non-uniqueness for $\boldsymbol \sigma$ comes from the existence of multiple extensions of interatomic potentials from the phase space to the shape space. Essentially, any given potential can be altered to an equivalent potential by adding a function of the Caley-Menger determinant (\cite{Admal-Tadmor2010}). Different extensions of a given potential energy can result in different force decompositions. Nevertheless, the difference due to any two pointwise stress tensors, resulting from different extensions for the interatomic potential energy, tend to zero as the volume of the domain over which these pointwise quantities are spatially averaged tends to infinity.

They (\cite{Admal-Tadmor2010}) also generalize their approach to possible non-straight interactions for systems with internal degrees of freedom. For such systems, balance of angular momentum is satisfied only through the presence of couples. The pointwise fields so constructed are not macroscopic continuum fields; at sufficient low temperature they are highly non-uniform, exhibiting a criss-cross pattern even when macroscopically the material is under uniform stress. To obtain true macroscopic quantities one should average over some spatial region surrounding the continuum point. For the case when spectroscopic data are converted to macroscopic quantities this area is dictated by the probe and its lengthscale. 

Based on the fact that the microscopic state of the system in never known to us and the only observables identified are the macroscopic fields as defined in continuum mechanics, Murdoch (\cite{Murdoch1982,Murdoch2003,Murdoch2007,Murdoch-Bedeaux1993,Murdoch-Bedeaux1994}) and Hardy (\cite{Hardy1982}) presented a procedure purely deterministic in nature. They defined continuum fields as direct spatial averages of the discrete equations of motion using weighted functions with compact support.  They proposed spatial averaging which can be followed by time averaging. Since under conditions of thermodynamic equilibrium ensemble averages can be replaced by time averages for an ergodic system, the approach of Murdoch (\cite{Murdoch1982,Murdoch2003,Murdoch2007,Murdoch-Bedeaux1993,Murdoch-Bedeaux1994}) and Hardy (\cite{Hardy1982}) can be obtained from the unified framework of Admal and Tadmor (\cite{Admal-Tadmor2010}). Following \cite{Admal-Tadmor2010}, the Murdoch (\cite{Murdoch1982,Murdoch2003,Murdoch2007,Murdoch-Bedeaux1994}) and Hardy (\cite{Hardy1982}) procedure can be interpreted as a probabilistic model constructed from the data obtained from a deterministic model. All in all, the unified framework of \cite{Admal-Tadmor2010} offers a very generic point of view regarding possible definitions of Cauchy's stress tensor. Under this perspective standard definitions such as the virial theorem, the Tsai traction (\cite{Tsai1979}) and the Hardy stress (\cite{Murdoch1982,Murdoch2003,Murdoch2007,Murdoch-Bedeaux1993,Murdoch-Bedeaux1994,Hardy1982}) can be seen under the same umbrella. 

An alternative approach to define the microscopic stress tensor which does not rely on the statement of balance of linear momentum is the approach adopted by Arroyo and co-workers (\cite{Torresetal2015,Torresetal2016}). The starting point in this geometric approach is the Doyle-Ericksen formula and what is achieved is the removal of the ambiguity in the definition of the microscopic stress in the presence of multibody interactions by naturally suggesting a canonical and physically motivated force decomposition into pairwise terms. This is a distinguished central force decomposition which is called covariant force decomposition (cCFD) and its derivation does not resort to Noll's lemma (\cite{Noll1955}). This covariant central force decomposition coincides with the common definition of the central force decomposition (CFD) for potentials with 4 or fewer body interactions. It is a close analog of CFD which replaces the partial differentiation of the interatomic potential by a covariant differentiation along the phase space. Since the shape space is an open subset of the phase space for $n\leq4$, cCFD and CFD coincide; however when $n>5$ cCFD resolves the ambiguity of the usual CFD.  

In this work, we adopt the framework of \cite{Admal-Tadmor2010} to examine to which stress tensor components the $E_{2g}$ and $A_{1g}$ vibration modes of one graphene (hexagonal) ring correspond. We start in Section 2 with a short reminder of the generic framework of \cite{Admal-Tadmor2010}. Having in mind that the $E_{2g}$ and $A_{1g}$ modes scatter an incoming photon inelastically and give rise to the so called Raman spectrum, we assume tacitly that a laser (a source of photons) necessary to activate these modes is present in our analysis. Then in Section 3.1 we focus on only one graphene ring and introduce the coordinate system with respect to which all of our calculations are done. The armchair direction is chosen to be along the x-axis, while the zig-zag direction along the y-axis. Calculations of this section are related with the formation stresses that make up the hexagonal ring out of the six carbon atoms. In Section 3.2 we study the $A_{1g}$ mode when graphene is at rest (namely, no applied external loading with the $A_{1g}$ mode active). We introduce this vibration as a suitable perturbation in the vector describing the bond direction and evaluate the stress components corresponding to this mode. We then do a similar analysis in Section 3.3 for the $E_{2g}$ mode: we use a suitable perturbation in the vector describing the bond direction and examine what happens to the stress tensor components. Our approach, though qualitative in nature, incorporates all the main physics and reveals that $E_{2g}$ and $A_{1g}$ vibration modes should be related with all types of stress components: shear as well as axial stress components. In order to bring our approach closer to more concrete results in Section 4 we evaluate the Hardy stress tensor that correspond to the $E_{2g}$ and $A_{1g}$ modes. Replacing phase average with time average (an assumption valid for ergodic systems under thermodynamic equilibrium) and choosing the Mie pair potential we are able to evaluate the Hardy stress tensor which supports the argument that all stress tensor components should be taken into account when converting data from Raman spectroscopic measurements when there is no applied external loading.

Our analysis extends to take into account an applied external tensile field, $z^{\text{appl}}$. When $z^{\text{appl}}$ is along the armchair direction, the higher order term of the axial $\sigma_{11}$ stress component of Cauchy's stress tensor is $(z^{\text{appl}})^2$, that of $\sigma_{12}$ is $z^{\text{appl}}$, while $z^{\text{appl}}$ is not present in the $\sigma_{22}$ stress component. We therefore infer that as $z^{\text{appl}}$ increases it is $\sigma_{11}$ which is the dominant stress component over all other stress components. When $z^{\text{appl}}$ is along the zig-zag direction, $\sigma_{22}$ stress component contain term $(z^{\text{appl}})^2$, $\sigma_{12}$ stress component contain term $z^{\text{appl}}$, while $\sigma_{11}$ component contain term $(\nu z^{\text{appl}})^2$, $\nu$ being graphene's Poisson ratio. Thus, it appears that as $z^{\text{appl}}$ increases it is the $\sigma_{22}$ stress component which is the dominant one over all other stress components. To generalize our analysis, we study tension applied at an arbitrary inclination described by angle $\theta$. In this case, the $\sigma_{11}$ component highest order term is $(z^{\text{appl}} \text{cos} \theta)^2$, that of $\sigma_{12}$ component is $(z^{\text{appl}})^2 \text{cos} \theta \text{sin} \theta$, while that of $\sigma_{22}$ is $(z^{\text{appl}} \text{sin} \theta)^2$. Thus, as $\theta$ approaches 45$^0$ the shear component increases reaching its maximum value at $\theta=45^0$ and at this value all stress components are of the same order. All these calculations are presented in Section 5: Section 5.1 studies the case when tension is along the armchair direction, Section 5.2 treats the case when tension is along the zig-zag direction, while the general case is treated in Section 5.3. 

There is a very rich and important solid state physics literature concerning the effect of externally applied strain (or stress) to graphene's Raman spectra (\cite{Ferrari2007,Ferrarietal2006,Ferrari-Basko2013,Ferrari-Robertson2001,Malardetal2009,Mohhiuddinetal2009,Sahataetal1988,Saitoetal2011,Ganesanetal1970}). These authors use the discrete equations of motion (Newton's second law) for the quasiharmonic approximation to measure the effect of externally applied axial stress/strain on the frequency of graphene's Raman spectra in similar trends with earlier approaches on polymer chains (\cite{Tashiroetal1990,Treloar1960}). Compared to these important studies (\cite{Ferrari2007,Ferrarietal2006,Ferrari-Basko2013,Ferrari-Robertson2001,Malardetal2009,Mohhiuddinetal2009,Sahataetal1988,Saitoetal2011,Ganesanetal1970}) our framework of Sections 3, 4 evaluates the stress tensor components which corresponds to frequency e.g. $\omega=1580 cm^{-1}$ for $E_{2g}$, namely to frequency which pertain to zero applied external strain. Since this frequency is not zero it should correspond to some kind of internal stresses; these are the stresses we evaluate in Section 3.2, 3.3. as well as in Section 4. In simple words, before the application of an external field the frequencies corresponding to $E_{2g}$ and $A_{1g}$ vibrations are not zero, thus produce internal stresses since these vibrational modes are active. Graphene is in equilibrium in such cases. 

The activation of $E_{2g}$ and $A_{1g}$ vibration modes produce internal stresses which are nevertheless equilibriated. The frequency of these modes at rest and at a non-zero temperature is non-zero, thus they correspond to non-zero stress components. It is these non-zero stress components that we evaluate in Sections 3.2, 3.3 and Section 4. It is in this sense that our approach should be seen as complementing that of \cite{Ferrari2007,Ferrarietal2006,Ferrari-Basko2013,Ferrari-Robertson2001,Malardetal2009,Mohhiuddinetal2009,Sahataetal1988,Saitoetal2011,Ganesanetal1970} in the sense that we evaluate the stress tensor components that correspond to the frequency of $E_{2g}$ and $A_{1g}$ vibration modes at rest (without the application of external stress or strain). 

When an external tensile field applies, we evaluate in Section 5 which stress components are the dominant one's depending on the geometry of the applied field. It turns out that when the external applied field is along the armchair direction the $\sigma_{11}$ stress component dominates in line with the approach of \cite{Sahataetal1988}. When the applied field is along the zig-zag direction, $\sigma_{22}$ stress component dominates, while for an arbitrary inclination of the applied tensile field all components, axial as well as shear, should be taken into account. Compared to the discrete approach of \cite{Ferrari2007,Ferrarietal2006,Ferrari-Basko2013,Ferrari-Robertson2001,Malardetal2009,Mohhiuddinetal2009,Sahataetal1988,Saitoetal2011,Ganesanetal1970} our framework follows that of \cite{Admal-Tadmor2010} which starts from the momentum equation which is a purely continuum concept, in contrast to the use of Newton's second law which corresponds to the purely discrete approach. To do a similar analysis and study the effect of stresses on frequency through the present framework, one should introduce plane progressive waves to measure the perturbations corresponding to the $E_{2g}$ and $A_{1g}$ modes. From the mathematical point of view such an approach is highly non-trivial since one then arrives to a system of integro-differential equations which complicates the analysis. Perhaps the recent work of Dayal (\cite{Dayal2017}) might serve as a guide in such an approach. All in all, our theoretical analysis provides qualitative results answering the question of which stress components are important when an external field is applied to an hexagonal ring and one can do Raman measurements of the G and D peak. 

\section{Generic expression for the stress tensor}

This section gives the main ingredients of the approach of Admal and Tadmor (\cite{Admal-Tadmor2010}) which expands on the previous framework of IKN (\cite{Irving-Kirkwood1950,Noll1955}). The pointwise Cauchy's stress tensor, $\boldsymbol \sigma$, in their analysis has two parts
\begin{equation}
\boldsymbol \sigma({\bf x}, t)=\boldsymbol \sigma^{\text{k}}({\bf x}, t)+\boldsymbol \sigma^{\text{v}}({\bf x}, t),
\end{equation}
where $\boldsymbol \sigma^{\text{k}}$ and $\boldsymbol \sigma^{\text{v}}$ are respectively the kinetic and the potential part of the pointwise stress tensor. The kinetic part is symmetric and expressed as
\begin{equation}
\boldsymbol \sigma^{\text{k}}({\bf x}, t)=-\sum_{\alpha}m_{\alpha} <({\bf v}^{\text{rel}} \otimes {\bf v}^{\text{rel}}) W|{\bf x}_{\alpha}={\bf x}>. 
\end{equation}
From the physical point of view it reflects the momentum flux associated with the vibrational kinetic energy portion of the internal energy. Regarding notation, $m_{\alpha}$ denotes the mass of particle $\alpha$, $W:\Gamma \times R^+\rightarrow R$ is the probability density function of class $C^1$ defined on all phase space $\Gamma$ for all time $t$ and $<F|{\bf x}_j={\bf x}>=\int_{\Gamma_j}Fd{\bf x}$, where $\Gamma_j$ is a $(6N-3)$ dimensional subspace of $\Gamma$ that arises upon discarding the spatial variable ${\bf x}_j$ belonging to $j$ and after performing the integration the free variable ${\bf x}_j$ is to be replaced by $\bf x$ (\cite{Noll1955}). Essentially, $<.>$ denotes phase averaging with the probability density function $W$ \footnote{A simple example of a stationary (time-independent) probability density function over the canonical ensemble is $W=\frac{1}{N! h^{3N}Z} e^{-\mathcal H/k_BT}$, $h$=Planck's constant, $N$=number of atoms, $k_B$=Boltzmann's constant, $T$=absolute temperature, $\mathcal H$ the Hamiltonian of the function and $Z$ the partition function (see \cite{Admal-Tadmor2010} for more information).}. Also ${\bf v}^{\text{rel}}={\bf v}_{\alpha}-{\bf v}$ is the velocity of particle $\alpha$ relative to the pointwise velocity field, ${\bf v}_{\alpha}$ being $\dot{\bf x}_{\alpha}$, while ${\bf v}({\bf x}, t)=\frac{\sum_{\alpha}m_{\alpha}<W{\bf v}_{\alpha}|{\bf x}_{\alpha}={\bf x}>}{{\sum_{\alpha}<W|{\bf x}_{\alpha}={\bf x}>}}$ is the pointwise velocity field. At very low temperatures velocities approach zero, thus one may assume that the kinetic contribution of stress equals to zero. This is an assumption that we adopt here since it does not alter the main outcomes of our analysis. Even if the kinetic part of the stress tensor is taken into account the main argument of this work remains valid. 

So, we focus on the potential part of the pointwise stress tensor which following \cite{Admal-Tadmor2010} takes the form
\begin{equation}
\boldsymbol \sigma^{\text{v}}({\bf x}, t)=\frac{1}{2} \int_{\mathcal R^3} \sum_{\alpha, \beta, \alpha \neq \beta} \frac{{\bf z} \otimes {\bf z}}{||\bf z||} \int_{s=0}^1<\frac{\partial V_{\text{int}}}{\partial r_{\alpha \beta}}W|{\bf x}_{\alpha}={\bf x}+s {\bf z}, {\bf x}_{\beta}={\bf x}-(1-s){\bf z}>ds d{\bf x},
\end{equation}
and gives at every point $\bf x$ the superposition of the expectation values of the force in all possible bonds passing through $\bf x$. The variable $\bf z$ selects a bond length and direction and the variable $s$ slides the bond through $\bf x$ from end to end. As noted in \cite{Admal-Tadmor2010,Admal-Tadmor2011,Torresetal2015,Torresetal2016} the pointwise stress tensor is not unique, since different extensions of a given potential energy, $V_{\text{int}}$, can result in different force decompositions. Nevertheless, such a  difference between any two pointwise stress tensors tend to zero as the volume of the domain over which these pointwise quantities are spatially averaged tends to infinity. So, the macroscopic stress tensor which is defined in the thermodynamic limit is always unique and is independent of the potential energy expression. 

In our work here we do not get into the analysis of non-uniqueness due to different extensions. Since our target is to examine to which stress tensor components the $E_{2g}$ and $A_{1g}$ vibration modes correspond, we assume that a potential $V_{\text{int}}$ exists and render the expression of the stress tensor unique in the thermodynamic limit. So, in the last relation $V_{\text{int}}=\hat{V}_{\text{int}}({\bf x}_1, {\bf x}_2, ..., {\bf x}_N)=V_{\text{int}}(r_{12},..., r_{(N-1)N})$ where $r_{\alpha \beta}=||{\bf x}_{\alpha}-{\bf x}_{\beta} ||$ and $V_{\text{int}}$ is the internal part of the potential energy, namely the interatomic potential assumed to be a $\hat{V}_{\text{int}}:R^{3N}\rightarrow R$ continuously differentiable function. Essentially, there is mapping $\Phi$ from the configuration space to the shape space, namely a mapping from the space of positions of particles to the space of distances of particles, $\Phi: R^{3N} \rightarrow S$, $S$ being the shape space which is a $(3N-6)$-dimensional manifold in $R^{N(N-1)/2}$, $\Phi:({\bf x}_1, {\bf x}_2,..., {\bf x}_N) \mapsto (r_{12},..., r_{(N-1)N})$, which allow us to change arguments on $V_{\text{int}}$. For the derivative of $V_{\text{int}}$ we have e.g. $\frac{d V_{int}}{d \zeta_{12}}({\bf s})=lim_{\lambda \rightarrow 0} \frac{V_{int}(r_{12}+\lambda, ..., r_{(N-1)N})-V_{int}(r_{12}, ..., r_{(N-1)N})}{\lambda}$. 

With respect to components eq. (3) can be written as 
\begin{equation}
\sigma_{ij}^{\text{v}}({\bf x}, t)=\frac{1}{2} \int_{\mathcal R^3} \sum_{\alpha, \beta, \alpha \neq \beta} \frac{{\bf z}_i \otimes {\bf z}_j}{||\bf z||} \int_{s=0}^1<\frac{\partial V_{\text{int}}}{\partial r_{\alpha \beta}}W|{\bf x}_{\alpha}={\bf x}+s {\bf z}, {\bf x}_{\beta}={\bf x}-(1-s){\bf z}>ds d{\bf x},
\end{equation}
which reveals that the nature of the stress components (axial $\sigma_{11}, \sigma_{22}$ or shear $\sigma_{12}, \sigma_{21}$) are related to the vector $\bf z$ which selects bond length and direction. This is a crucial observation for the following analysis: it is the term ${\bf z}_i \otimes {\bf z}_j$ that characterizes the nature of the stress tensor components. We also note that since graphene is a genuinely 2D material components $\sigma_{33}, \sigma_{32}, \sigma_{31}$ of the stress tensor have no physical meaning.  

The pointwise stress field $\boldsymbol \sigma^{\text{v}}(\bf x, t)$ so defined is not a macroscopic quantity. At sufficient low temperatures it is highly non uniform exhibiting a criss-cross pattern with higher stresses along bond directions even in cases when macroscopically the material is under uniform stress. To obtain the macroscopic field $f_{\omega}({\bf x}, t)$ from a pointwise field $f({\bf x}, t)$ one should average over some spatial region surrounding the continuum point (\cite{Admal-Tadmor2010})
\begin{equation}
f_{\omega}({\bf x}, t)=\int_{\mathcal R^3} \omega({\bf y}-{\bf x}) f({\bf y}, t) d{\bf y}. 
\end{equation}
Function $\omega({\bf r})$ is a weighting function representing the properties of the probe and its lengthscale. One possible choice is to take $\omega({\bf r})$ to be a $R^+$ valued function with compact support such that $\omega({\bf r})=0$, for $||{\bf r}||>\lambda$, where $\lambda$ is the lengthscale connected with the probe of the Raman measurement. A Raman microprobe with a laser excitation length at 514 nm can scan a specimen of regions as small as approximately 500nm; it is this value that gives the region over which the pointwise quantity should be averaged. From the theoretical point of view the work of \cite{Ulzetal2013} put forth a hypothesis to obtain a lower bound for the size of the spatial averaging volume. Our analysis here is motivated from experiments where the spatial averaging volume is dictated by the Raman spectrometer and its resolution. 

So, the true macroscopic Cauchy stress tensor is written as 
\begin{equation}
{\boldsymbol \sigma}^{\text{v}}_{\omega}({\bf x}, t)=\int_{\mathcal R^3} \omega({\bf y}-{\bf x}) {\boldsymbol \sigma}^{\text{v}}({\bf y}, t) d{\bf y}
\end{equation}
and by using the expression of eq. (4) for $\boldsymbol \sigma^{\text{v}}$ it can be seen that it is the vector $\bf z$ which determines the nature of the components of the stress tensor, $\boldsymbol \sigma^{\text{v}}_{\omega}$. In line with Cauchy's standard argument the traction vector is defined as
\begin{eqnarray}
&&{\bf t}({\bf x}, {\bf n}; t)={\boldsymbol \sigma^{\text{v}}_{\omega}} {\bf n}= \nonumber\\ 
&&\int_{\mathcal R^3} \frac{1}{2} \sum_{\alpha, \beta, \alpha \neq \beta} \frac{{\bf z} \otimes {\bf z}}{||\bf z||} \int_{s=0}^1<\frac{\partial V_{\text{int}}}{\partial r_{\alpha \beta}}W|{\bf x}_{\alpha}={\bf x}+s {\bf z}, {\bf x}_{\beta}={\bf x}-(1-s){\bf z}> \times ({\bf z} {\bf n})ds d{\bf x}
\end{eqnarray}
and gives a measure of the force per unit area of all the bonds that cross the surface. Certainly, it is the traction corresponding to the potential part of the stress tensor. 

In the subsequent sections we focus on eq. (4) namely on the expression of the potential part of the stress tensor and by making appropriate perturbation hypothesis we introduce the $E_{2g}$ and $A_{1g}$ vibration modes into the expression of the stress tensor components. Namely, we evaluate to which stress components each of these vibration modes correspond. This can be straightforwardly extended to a true macroscopic quantity by averaging out using eq. (6), since such a procedure does not alter qualitative the results.  

\section{Stress interpretation of $E_{2g}$ and $A_{1g}$ modes.}
 
In Section 3.1 we focus on only one graphene ring and assume that we are confined to very low temperatures such that the kinetic term of eq. (2) is absent \footnote{Essentially, this analysis is carried out for temperatures near 0K, but the qualitative outcomes remain the same even for higher temperatures.}. By placing a Cartesian coordinate system we are in a position to find the components of all carbon atoms of the ring at rest. We choose to place the armchair direction along the x-axis while the zig-zag direction is along y-axis and we use carbon atom components to find the stress tensor components for the hexagonal ring at rest. These are essentially the formation stresses, namely the stresses necessary to keep the six carbon atoms at the hexagonal form: it is the formation energy of a graphene ring. To find to which stress tensor components the $E_{2g}$ and $A_{1g}$ modes correspond, we introduce appropriate perturbations into the expression for the stress tensor. These perturbations are introduced through term $\bf z$ in eq. (4). This is done in sections 3.2 and 3.3: in section 3.2 we focus on the $A_{1g}$ mode while in section 3.3 we study the $E_{2g}$ mode. The main outcome is that Raman spectra corresponding to $E_{2g}$ and $A_{1g}$ vibration modes should be related with all types of stress components: axial as well as shear. Since graphene is assumed to be in equilibrium, which correspond to frequency of e.g. $\omega=1580 cm^{-1}$ for $E_{2g}$ \footnote{This value for the frequency if for room temperature, while the analysis carried out here is for 0K; nevertheless it gives a qualitative value of the frequency.}, calculations of Section 3.2 and 3.3 measure stresses for this frequency; namely, for graphene at rest without any applied mechanical field.  

\subsection{Internal stresses}

We focus on only one hexagonal ring and use the Axy coordinate system as seen in Figure 1. 
\begin{figure}[!htb]
\centering
\includegraphics[width=80mm]{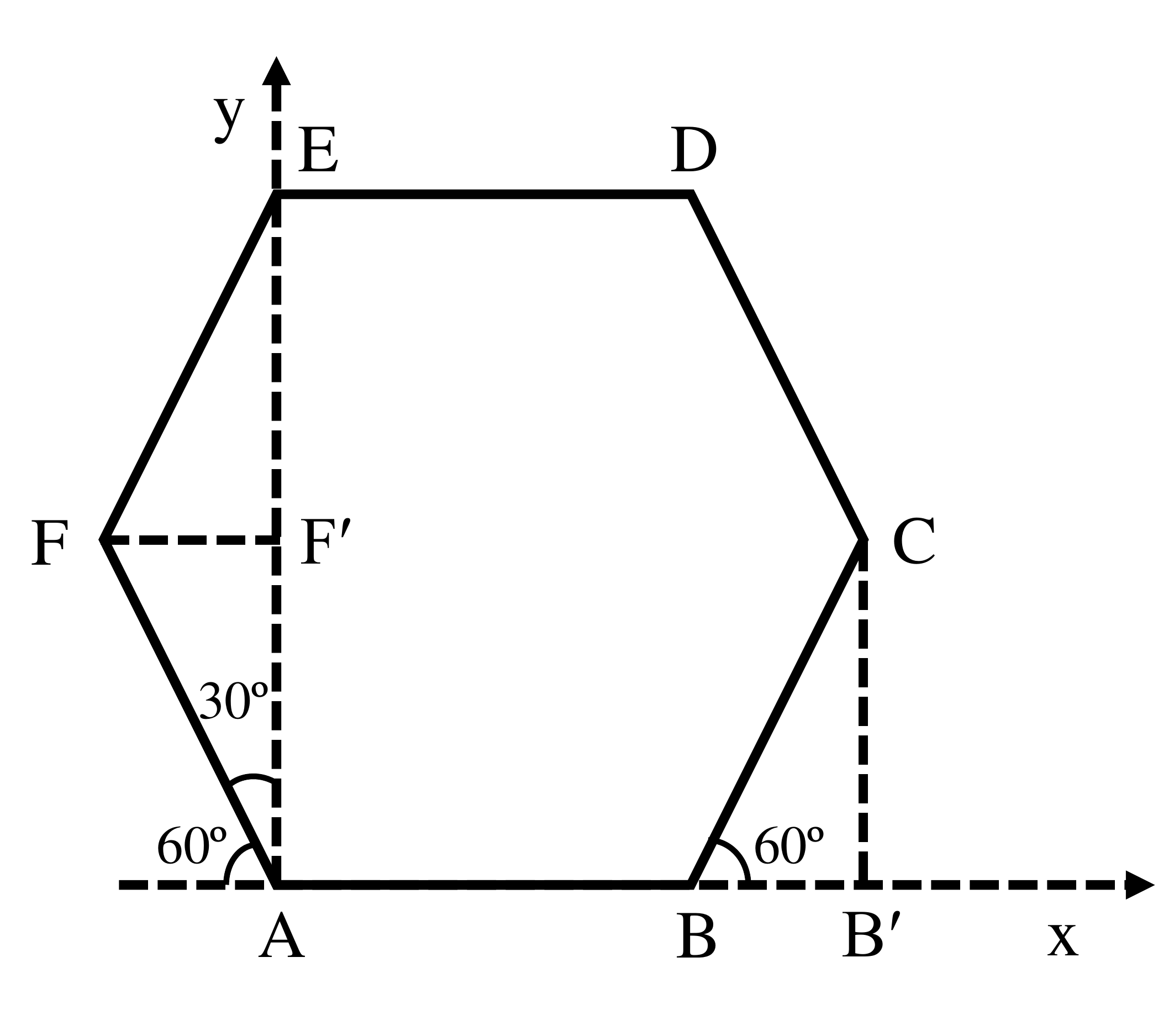}
\caption{One graphene ring at rest and the Axy coordinate system used for evaluating eqs. (8-12).}
\end{figure}
The armchair direction is placed along the x-axis and the zig-zag direction along the y-axis, without any restrictions in the generality. With respect to this coordinate system the components of the positions occupied by carbon atoms have coordinates $A(0, 0)$, $B(l, 0)$, $C(l+lcos60^0, lcos30^0)$, $D(l, 2lcos30^0)$, $E(0, 2lcos30^0)$, $F(lcos60^0, -lcos30^0)$ where the interatomic distance at rest is approximately $l=1.42\AA$. Since the expression of eq. (4) contain the bond vector $\bf z$ connecting two carbon atoms we introduce two unit vectors as basis for the system above: these are vectors ${\bf z}_1, {\bf z}_2$ which are parallel to the Ax and Ay axis, respectively. So, for each bond vector connecting two carbon atoms we have
\begin{eqnarray}
{\bf z}_{AB}&=& \alpha_{AB}{\bf z}_1, \alpha_{AB}=|AB|=1,42\AA, |{\bf z}_1|=1\AA, \\
{\bf z}_{AC}&=& \alpha_{AC}{\bf z}_1+\beta_{AC}{\bf z}_2, \alpha_{AC}=|AB|+|BC|cos60=2,13\AA, \nonumber\\
&& \ \ \ \ \ \ \ \ \ \ \ \ \ \ \ \ \ \ \ \ \ \  \beta_{AC}=|AF'|=|AF|sin60=1,229\AA, \\
{\bf z}_{AD}&=& \alpha_{AD}{\bf z}_1+\beta_{AD}{\bf z}_2, \alpha_{AD}=\alpha_{AC}=2,13\AA, \beta_{AD}=2\beta_{AC}=2,258\AA, \\
{\bf z}_{AE}&=& \alpha_{AE}{\bf z}_1+\beta_{AE}{\bf z}_2, \alpha_{AE}=0, \beta_{AE}=2\beta_{AC}=2,458\AA, \\
{\bf z}_{AF}&=& \alpha_{AF}{\bf z}_1+\beta_{AF}{\bf z}_2, \alpha_{AF}=-|AF|sin60=-1,229\AA, \beta_{AF}=\beta_{AC}=1,229\AA.
\end{eqnarray}

We note that henceforth we drop the $\AA$ designation, but we stress that length measure quantities are used in our tensor products and expressions of $\bf z$. From the above equations one evaluates tensor products
\begin{eqnarray}
{\bf z}_{AB} \otimes {\bf z}_{AB}&=&2,0164 {\bf z}_1 \otimes {\bf z}_1, \\
{\bf z}_{AC} \otimes {\bf z}_{AC}&=&4,5369 {\bf z}_1 \otimes {\bf z}_1+2,3824{\bf z}_1 \otimes {\bf z}_2+2,3824{\bf z}_2 \otimes {\bf z}_1+1,5104{\bf z}_2 \otimes {\bf z}_2, \\
{\bf z}_{AD} \otimes {\bf z}_{AD}&=&4,5369 {\bf z}_1 \otimes {\bf z}_1+4,8095{\bf z}_1 \otimes {\bf z}_2+4,8095{\bf z}_2 \otimes {\bf z}_1+5,0985{\bf z}_2 \otimes {\bf z}_2, \\
{\bf z}_{AE} \otimes {\bf z}_{AE}&=&6,0417 {\bf z}_2 \otimes {\bf z}_2, \\
{\bf z}_{AF} \otimes {\bf z}_{AF}&=&1,5104 {\bf z}_1 \otimes {\bf z}_1-1,5104{\bf z}_1 \otimes {\bf z}_2-1,5104{\bf z}_2 \otimes {\bf z}_1+1,5104{\bf z}_2 \otimes {\bf z}_2.
\end{eqnarray}
Putting these expressions to eq. (4) one finds for the formation stresses at point A
\begin{eqnarray}
\boldsymbol \sigma^{\text{v}}_{\text{int}}&=&\frac{1}{2} \int_{\mathcal R^3} \frac{12,6452 {\bf z}_1 \otimes {\bf z}_1+5,6825 ({\bf z}_1 \otimes {\bf z}_2+{\bf z}_2 \otimes {\bf z}_1)+14,1603{\bf z}_2 \otimes {\bf z}_2}{17,437} \nonumber\\
&&<\frac{\partial V_{\text{int}}}{\partial r_{\alpha \beta}}W|{\bf x}_{\alpha}={\bf x}, {\bf x}_{\beta}={\bf x}-{\bf z}> d{\bf x}.
\end{eqnarray}

Averaging out using eq. (6) one can calculate the potential part of the stresses necessary in order to be able to form a hexagonal ring out of six carbon atoms (not related with respect to each other). As can be seen from the componential expression of $\boldsymbol \sigma^{\text{v}}$ in eq. (4), it is the term $\bf z$ that characterizes the stress component. For our analysis above ${\bf z}_1 \otimes {\bf z}_1$, ${\bf z}_2 \otimes {\bf z}_2$ mean axial components along the Ax and Ay directions, namely $\sigma_{11}, \sigma_{22}$ respectively, while ${\bf z}_1 \otimes {\bf z}_2 + {\bf z}_2 \otimes {\bf z}_1$ mean shear stress component, $\sigma_{12}$. It is also true that since the phase space is not known one cannot integrate but nevertheless he can have a qualitative measure of which kind of bonds are related with pure axial or shear components. 

We also mention that the specific choice of the potential $V_{\text{int}}$ describing the interaction can also change the outcome. For example, for carbon atom at position A a potential that takes into account only first neighbors it only takes account of carbon atoms at B and F. A potential that takes into account second neighbors additionally see atoms C and E, etc. What remains to be said is that it matters which point $\bf x$ of the hexagonal ring one wants to study. Eq. (18) gives the stress tensor at point A when only straight bond interactions between carbon atoms are taken into account. One may find similar expressions for all other positions occupied by carbon atoms. Then the stress tensor for points lying in the line between two carbon atoms is found by adjusting variable $s$ accordingly. Also, non-straight bonds interactions could have been taken into account following the framework of \cite{Admal-Tadmor2010}. For a point within the hexagonal ring that does not lie in a line joining two carbon atoms, stresses evaluated from the above analysis are zero. Only when non-straight interactions are taken into account, stresses are zero for such points. 

So, all in all, in this subsection we give the expression of carbon distances for only one hexagonal ring at rest; these are eqs. (8-12). Using these expressions to eqs. (4, 6) one may evaluate the macroscopic stresses necessary to form the hexagonal ring out of six carbon atoms. Essentially, we speak about the formation stresses that keep graphene in its hexagonal form and consist of axial $\sigma_{11}, \sigma_{22}$ (stemming from terms ${\bf z}_1 \otimes {\bf z}_1$, ${\bf z}_2 \otimes {\bf z}_2$, respectively) as well as shear stress $\sigma_{12}$ component (stemming from terms ${\bf z}_1 \otimes {\bf z}_2 + {\bf z}_2 \otimes {\bf z}_1$). 

\subsection{$A_{1g}$ mode-D peak}

To evaluate to which stress tensor components the D-peak corresponds we assume appropriate perturbations corresponding to the breathing mode of the $A_{1g}$ mode that can be seen on Figure 2. 
\begin{figure}[!htb]
\centering
\includegraphics[width=80mm]{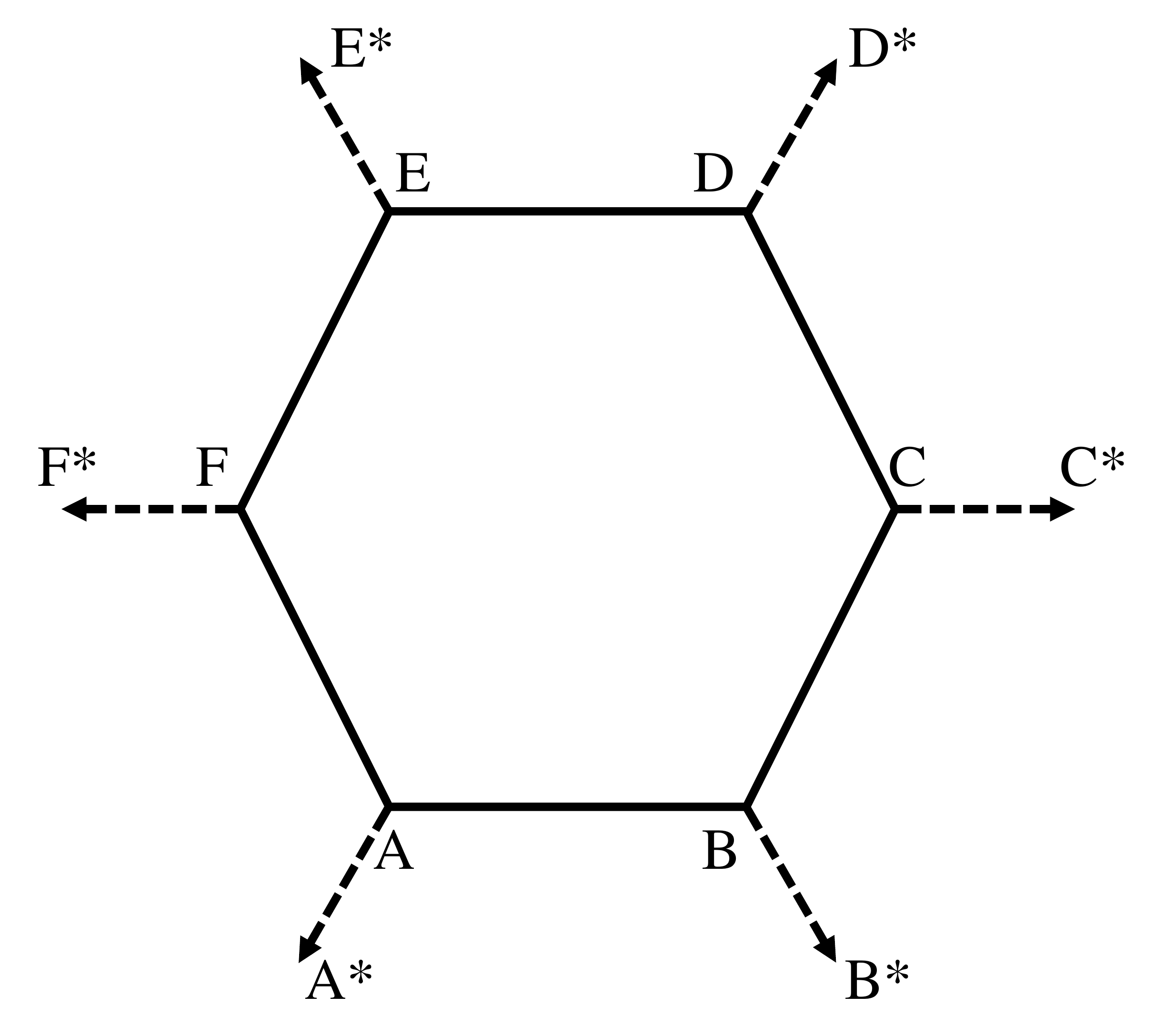}
\caption{The breathing mode corresponding to the D-peak.}
\end{figure}
Perturbed quantities are designated by a superposed asterisk, $*$. Having in mind that the $A_{1g}$ scatter inelastically an incoming photon and give rise to the so-called Raman spectra, we assume that a laser excites the $A_{1g}$ mode and we examine to what stress tensor components this mode corresponds. Our outcome is independent of the external loading and what is only needed is the $A_{1g}$ mode to be activated; namely, graphene is at rest. We find after some straightforward calculations the perturbed quantities as
\begin{eqnarray}
{\bf z}^*_{AB}&=&{\bf z}_{AB}+\epsilon \delta u({\bf z}_1+0,866{\bf z}_2) , \\
{\bf z}^*_{AC}&=&{\bf z}_{AC}+\epsilon \delta u(1,5{\bf z}_1+0,866{\bf z}_2), \\
{\bf z}^*_{AD}&=&{\bf z}_{AD}+\epsilon \delta u({\bf z}_1+0,732{\bf z}_2), \\
{\bf z}^*_{AE}&=&{\bf z}_{AE}+\epsilon \delta u({\bf z}_1+0,732{\bf z}_2), \\
{\bf z}^*_{AF}&=&{\bf z}_{AF}+\epsilon \delta u(1,5{\bf z}_1+0,866{\bf z}_2),
\end{eqnarray}
and in all the above $\delta u$ is the length of the small perturbations ($\delta u=BB^*=AA^*=...$) assumed to be equal for each carbon atom. Also $\epsilon$ is a constant that is small enough to allow us to neglect the second and higher powers of $\epsilon$ compared with $\epsilon$ as is done in common plane progressive wave approaches (\cite{Sfyris2011}).

Putting these perturbed quantities back in the stress tensor of eq. (4) one observes that it is the term ${\bf z} \otimes {\bf z}$ which characterizes the nature of the stress tensor. For this quantity and for point A taking into account only straight bond interactions with all atoms of the hexagon we evaluate
\begin{equation}
{\bf z}^* \otimes {\bf z}^*={\bf z}^*_{AB} \otimes {\bf z}^*_{AB}+{\bf z}^*_{AC} \otimes {\bf z}^*_{AC}+{\bf z}^*_{AD} \otimes {\bf z}^*_{AD}+{\bf z}^*_{AE} \otimes {\bf z}^*_{AE}+{\bf z}^*_{AF} \otimes {\bf z}^*_{AF}. 
\end{equation}
Working on each term separately we have for the first term 
\begin{equation}
{\bf z}^*_{AB} \otimes {\bf z}^*_{AB}=({\bf z}_{AB}+\epsilon \delta u({\bf z}_1+0,866{\bf z}_2)) \otimes ({\bf z}_{AB}+\epsilon \delta u({\bf z}_1+0,866{\bf z}_2),
\end{equation}
which when terms of order of $\epsilon^2$ are neglected render after using eq. (8) we find
\begin{equation}
{\bf z}^*_{AB} \otimes {\bf z}^*_{AB}={\bf z}_{AB} \otimes {\bf z}_{AB}+\epsilon \delta u (2,84{\bf z}_1 \otimes {\bf z}_1+2,424({\bf z}_2 \otimes {\bf z}_1+{\bf z}_1 \otimes {\bf z}_2)). 
\end{equation}

Using similar reasoning for the second term on the right hand of eq. (24) we find after using eq. (9)
\begin{equation}
{\bf z}^*_{AC} \otimes {\bf z}^*_{AC}={\bf z}_{AC} \otimes {\bf z}_{AC}+\epsilon \delta u (6,51 {\bf z}_1 \otimes {\bf z}_1+5,487({\bf z}_1 \otimes {\bf z}_2 +{\bf z}_2 \otimes {\bf z}_1)+4,256{\bf z}_2 \otimes {\bf z}_2). 
\end{equation}
For the third term on the right hand side of eq. (24) we have when we use eq. (10)
\begin{equation}
{\bf z}^*_{AD} \otimes {\bf z}^*_{AD}={\bf z}_{AD} \otimes {\bf z}_{AD}+\epsilon \delta u (4,26 {\bf z}_1 \otimes {\bf z}_1+6,147 ({\bf z}_1 \otimes {\bf z}_2 + {\bf z}_2 \otimes {\bf z}_1)+8,514 {\bf z}_1 \otimes {\bf z}_2). 
\end{equation}
For the fourth term we have after using eq. (11)
\begin{equation}
{\bf z}^*_{AE} \otimes {\bf z}^*_{AE}={\bf z}_{AE} \otimes {\bf z}_{AE}+\epsilon \delta u (2,458 ({\bf z}_2 \otimes {\bf z}_1+ {\bf z}_1 \otimes {\bf z}_2) + 8,514 {\bf z}_2 \otimes {\bf z}_2). 
\end{equation}
The last term on the right hand side of eq. (24) give using eq. (12)
\begin{equation}
{\bf z}^*_{AF} \otimes {\bf z}^*_{AF}={\bf z}_{AF} \otimes {\bf z}_{AF}+\epsilon \delta u (-2,907 {\bf z}_1 \otimes {\bf z}_1+0,779( {\bf z}_1 \otimes {\bf z}_2 + {\bf z}_2 \otimes {\bf z}_1) + 2,128 {\bf z}_2 \otimes {\bf z}_2). 
\end{equation}

So, collectively the stress tensor expression stemming from such perturbations takes the form 
\begin{eqnarray}
\boldsymbol \sigma^{\text{v}}({\bf x}, t)&=&\boldsymbol \sigma^{\text{v}}_{\text{int}}+\frac{1}{2} \epsilon \int_{\mathcal R^3} \delta u \frac{10,703 {\bf z}_1 \otimes {\bf z}_1+20,975 ({\bf z}_1 \otimes {\bf z}_2+{\bf z}_2 \otimes {\bf z}_1)+15,412{\bf z}_2 \otimes {\bf z}_2}{6,142} \nonumber\\
&&<\frac{\partial V_{\text{int}}}{\partial r_{\alpha \beta}}W|{\bf x}_{\alpha}={\bf x}, {\bf x}_{\beta}={\bf x}-{\bf z}> d{\bf x},
\end{eqnarray}
where $\boldsymbol \sigma^{\text{v}}_{\text{int}}$ are the internal stresses produced from the unperturbed quantities, eq. (18). The other part of the stress tensor, namely 
\begin{eqnarray}
&&\frac{1}{2} \int_{\mathcal R^3} \delta u \frac{10,703 {\bf z}_1 \otimes {\bf z}_1+20,975 ({\bf z}_1 \otimes {\bf z}_2+{\bf z}_2 \otimes {\bf z}_1)+15,412{\bf z}_2 \otimes {\bf z}_2}{6,142}\nonumber\\ 
&&<\frac{\partial V_{\text{int}}}{\partial r_{\alpha \beta}}W|{\bf x}_{\alpha}={\bf x}, {\bf x}_{\beta}={\bf x}-{\bf z}> d{\bf x},
\end{eqnarray}
is the stress tensor related with the perturbation of the D-peak mode. Essentially, it has four terms and it can be written in matrix form as
\begin{equation}
\begin{bmatrix}
    10,703 & 20,975  \\
    20,975 & 15,412 
\end{bmatrix} {\bf z}_i \otimes {\bf z}_j
\end{equation}
This means that term $10,703{\bf z}_1 \otimes {\bf z}_1$ is related with the axial stress $\sigma_{11}$ component along the x-axis while term $15,412{\bf z}_2 \otimes {\bf z}_2$ is the axial stress $\sigma_{22}$ along the y-axis produced from the perturbation of the D-peak. Terms $20,975 ({\bf z}_1 \otimes {\bf z}_2+{\bf z}_2 \otimes {\bf z}_1)$ is the shear stress component $\sigma_{12}$ produced from the D-peak type of perturbation. Certainly, to these terms one should incorporate the rest part of the stress tensor which concerns the phase space integration as well as the spatial average of eq. (6). Nevertheless, these type of integrations do not alter the character of the stress tensor corresponding to the perturbations of the D-peak neither do they eliminate some of these terms, in general. Thus, from the above analysis one infers that the Cauchy stress tensor related with all stress components: axial as well as shear stress components should be produced from a breathing mode of the type of the D-peak as far as the $A_{1g}$ mode is activated. 

Eq. (32) gives the stress tensor for point A and by similar calculations one may evaluate the stress tensor at every point occupied by a carbon atom. For points belonging to lines connecting two carbon atoms one should adjust the quantity $s$ accordingly. For points not belonging to a straight line joining particles one should use the non-straight interactions introduced by \cite{Admal-Tadmor2010} in order to obtain a non-trivial value for the stress tensor. In such a case perhaps a non-symmetric tensor should come out of the analysis, since as noted in \cite{Admal-Tadmor2010}, these kind of interactions might suit to a body with internal structure characterized by a non-symmetric stress tensor. To all these considerations one should add the spatial average step in order the so defined stress tensor to be a macroscopic quantity. Nevertheless, the spatial average does not alter the character of the stress tensor which should contain shear as well as all kind of axial components. All in all, our analysis shows that the breathing mode that corresponds to the D-peak in a Raman diagram should be related with a full stress tensor in the sense that axial and shear components are produced when this mode is activated.

\subsection{$E_{2g}$ mode: the G-peak}

Schematically the perturbations related with the $E_{2g}$ mode can be seen in Figure 3. 
\begin{figure}[!htb]
\centering
\includegraphics[width=80mm]{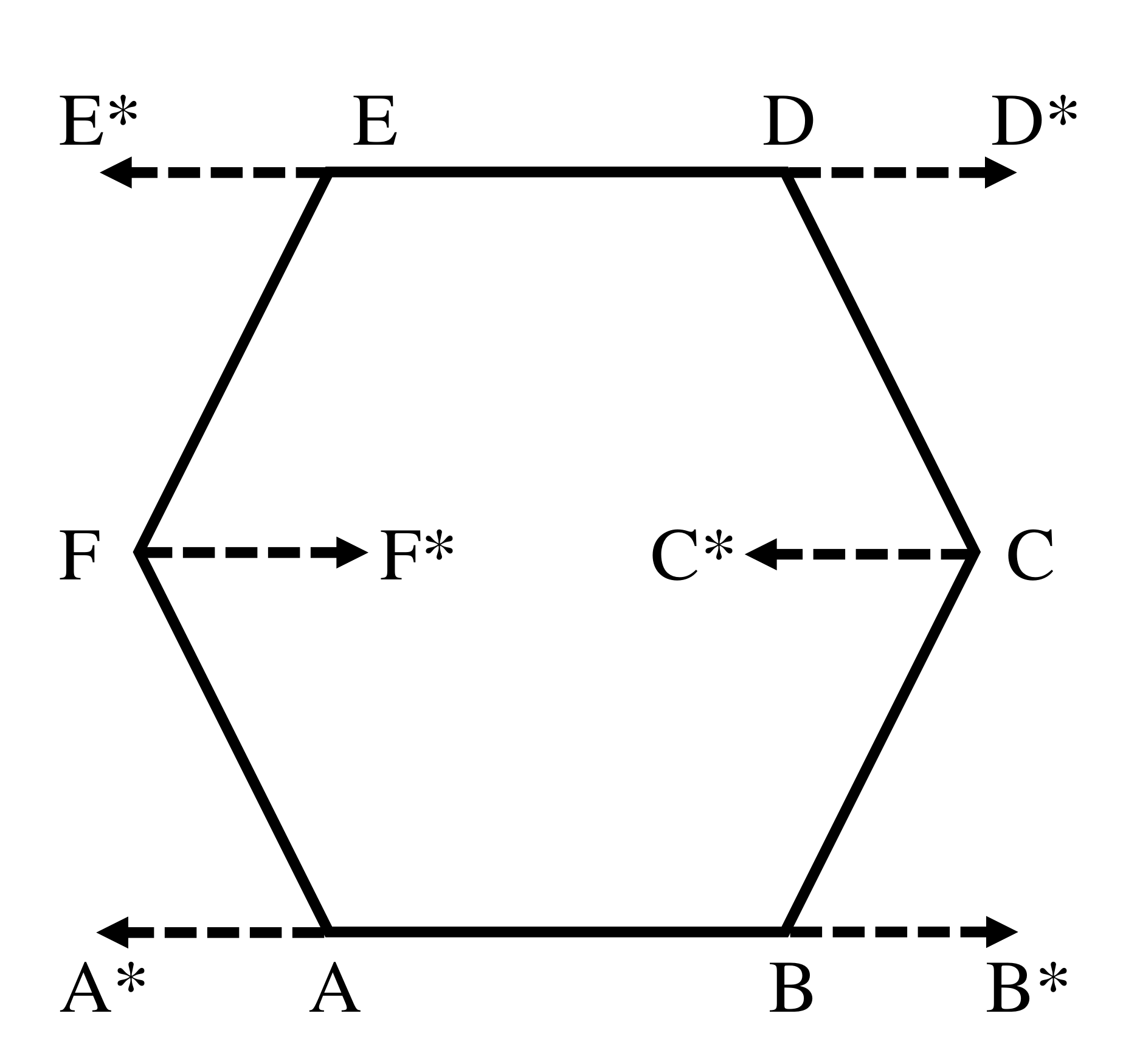}
\caption{Perturbations corresponding to the $E_{2g}$ mode. }
\end{figure}
As for the $A_{1g}$ mode in Section 3.2, we here assume that a laser excites the graphene ring such that the $E_{2g}$ mode activate and we examine to which stress tensor components this activation corresponds. From the mathematical point of view these perturbations can be written as
\begin{eqnarray}
{\bf z}^*_{AB}&=&{\bf z}_{AB}+2 \epsilon \delta u {\bf z}_1, \\
{\bf z}^*_{AC}&=&{\bf z}_{AC}+2 \epsilon \delta u {\bf z}_1, \\
{\bf z}^*_{AD}&=&{\bf z}_{AD}+2 \epsilon \delta u {\bf z}_1, \\
{\bf z}^*_{AE}&=&{\bf z}_{AF}+2 \epsilon \delta u {\bf z}_1, \\
{\bf z}^*_{AF}&=&{\bf z}_{AE}+2 \epsilon \delta u {\bf z}_1. 
\end{eqnarray}
We note that in eq. (34-38) perturbations are taken in their absolute values (as is the case in eqs. (19-23) as well). This observation has physical meaning, for if it wasn't true then perturbation $BB^*$ would cancel $AA^*$ out so no stress would have been produced which is clearly not physically meaningful. So, after using eqs. (8-12) we can find
\begin{eqnarray}
{\bf z}^*_{AB} \otimes {\bf z}^*_{AB}&=&{\bf z}_{AB} \otimes {\bf z}_{AB}+5,68 \epsilon \delta u {\bf z}_1 \otimes {\bf z}_1, \\
{\bf z}^*_{AC} \otimes {\bf z}^*_{AC}&=&{\bf z}_{AC} \otimes {\bf z}_{AC}+\epsilon \delta u ( 8,52 {\bf z}_1 \otimes {\bf z}_1+2,458 {\bf z}_2 \otimes {\bf z}_1+2,458 {\bf z}_1 \otimes {\bf z}_2), \\
{\bf z}^*_{AD} \otimes {\bf z}^*_{AD}&=&{\bf z}_{AD} \otimes {\bf z}_{AD}+ \epsilon \delta u ( 8,52 {\bf z}_1 \otimes {\bf z}_1+4,916 {\bf z}_2 \otimes {\bf z}_1+4,916 {\bf z}_1 \otimes {\bf z}_2),  \\
{\bf z}^*_{AE} \otimes {\bf z}^*_{AE}&=&{\bf z}_{AE} \otimes {\bf z}_{AE}+ \epsilon \delta u (4,916 {\bf z}_1 \otimes {\bf z}_2+4,916 {\bf z}_2 \otimes {\bf z}_1), \\
{\bf z}^*_{AF} \otimes {\bf z}^*_{AF}&=&{\bf z}_{AF} \otimes {\bf z}_{AF}+  \epsilon \delta u ( -4,916 {\bf z}_1 \otimes {\bf z}_1-4,916 {\bf z}_2 \otimes {\bf z}_1-4,916 {\bf z}_1 \otimes {\bf z}_2). 
\end{eqnarray}

Thus, for the stress tensor components we have for point A
\begin{eqnarray}
\boldsymbol \sigma^{\text{v}}({\bf x}, t)&=&\boldsymbol \sigma^{\text{v}}_{\text{int}}+\frac{1}{2} \epsilon \int_{\mathcal R^3} \delta u \frac{17,804 {\bf z}_1 \otimes {\bf z}_1+15,046 ({\bf z}_1 \otimes {\bf z}_2+ {\bf z}_2 \otimes {\bf z}_1)}{27,744} \nonumber\\
&&<\frac{\partial V_{\text{int}}}{\partial r_{\alpha \beta}}W|{\bf x}_{\alpha}={\bf x}, {\bf x}_{\beta}={\bf x}-{\bf z}> d{\bf x},
\end{eqnarray}
where $\boldsymbol \sigma^{\text{v}}_{\text{int}}$ are the internal stresses as in eq. (18). The stress tensor related purely with the $E_{2g}$ mode is 
\begin{eqnarray}
&&\frac{1}{2} \int_{\mathcal R^3} \delta u \frac{17,804 {\bf z}_1 \otimes {\bf z}_1+15,046 ({\bf z}_1 \otimes {\bf z}_2+{\bf z}_2 \otimes {\bf z}_1)}{27,744}\nonumber\\ 
&&<\frac{\partial V_{\text{int}}}{\partial r_{\alpha \beta}}W|{\bf x}_{\alpha}={\bf x}, {\bf x}_{\beta}={\bf x}-{\bf z}> d{\bf x}
\end{eqnarray}
and clearly it is related with axial as well as shear stress components. Axial component $\sigma_{11}$ is related with term ${\bf z}_1 \otimes {\bf z}_1$ while shear stress $\sigma_{12}$ component is related with term ${\bf z}_1 \otimes {\bf z}_2+ {\bf z}_2 \otimes {\bf z}_1$. It appears that axial components along the y-direction do not appear, a reasonable result since all perturbations are along the y-axis. So, for the $E_{2g}$ mode a similar outcome as the one for $A_{1g}$ is valid which can be summarized as: the $E_{2g}$ mode when activated give rise to axial (along the y-direction solely) as well as shear stress components. Essentially, stress tensor components of eq. (45) correspond to frequency $\omega=1580 cm^{-1}$ for $E_{2g}$ in a Raman frequency diagram, namely for graphene at rest without the application of any external mechanical loading.  

\section{Hardy stress}

As mentioned by Murdoch (\cite{Murdoch1982,Murdoch2003,Murdoch2007,Murdoch-Bedeaux1993,Murdoch-Bedeaux1994}) and Hardy (\cite{Hardy1982}) there is in general a lack of knowledge of the ensemble of the system, thus phase averaging is practically useless in real applications. For that reason Murdoch (\cite{Murdoch1982,Murdoch2003,Murdoch2007,Murdoch-Bedeaux1993,Murdoch-Bedeaux1994}) proposed that a time average should be used instead of ensemble average for the evaluation of macroscopic quantities. This purely deterministic approach smears a discrete system to form a continuum using a weighting function $\omega$ and can be interpreted as a probabilistic model constructed from data obtained from a deterministic model (e.g. a molecular dynamics system). In the steps followed in this so-called Murdoch-Hardy procedure (\cite{Admal-Tadmor2010}) one starts by forming a continuum system by smearing the discrete system out and then introducing a non-local constitutive law for the continuum consistent with the discrete version of force balance. For each constitutive law a stress tensor can be defined which satisfies the equation of momentum. 

The Hardy stress (\cite{Hardy1982,Admal-Tadmor2010}) gives an instantaneous definition for a pair potential and can be derived from the generic approach of Admal and Tadmor (\cite{Admal-Tadmor2010}), as well as from the Murdoch-Hardy procedure (\cite{Murdoch1982,Murdoch2003,Murdoch2007,Murdoch-Bedeaux1993,Murdoch-Bedeaux1994,Hardy1982}). Essentially, replacing ensemble averages with time averages allows one to obtain the Hardy stress from the generic probabilistic framework of \cite{Admal-Tadmor2010}. This is feasible for an ergodic system under conditions of thermodynamic equilibrium. So, the Hardy stress tensor is valid under non-equilibrium conditions in cases the system is in local thermodynamic equilibrium at every instant of time and for every point. This is feasible when the microscopic equilibriation time scale $\tau$ and the microscopic time $t$ are clearly separable in the sense that $\tau$ is sufficiently small such that macroscopic observables do not vary appreciably over it. 

The dynamic part of the Hardy stress tensor, which is symmetric, read then (\cite{Admal-Tadmor2010})
\begin{equation}
\boldsymbol \sigma^{\text{v}}_{\omega}({\bf x}, t)=\frac{1}{2 \tau} \sum_{\alpha, \beta, \alpha \neq \beta} \int_t^{t+\tau}[-{\bf f}_{\alpha \beta} \otimes ({\bf x}_{\alpha}-{\bf x}_{\beta}) b({\bf x}; {\bf x}_{\alpha}, {\bf x}_{\beta})] dt,  
\end{equation}
where
\begin{equation}
b({\bf x}; {\bf v}, {\bf u})=\int_{s=0}^1 \omega((1-s){\bf v}+s{\bf u}-{\bf x}) ds
\end{equation}
is the bond function, namely the integral of the weighting function, $\omega$, centered at $\bf x$ along the lines connecting points $\bf v$ and $\bf u$. The central force term ${\bf f}_{\alpha \beta}$ is expressed as 
\begin{equation}
{\bf f}_{\alpha \beta}({\bf u})=\frac{\partial V_{int}}{\partial \zeta_{\alpha \beta}} \frac{{\bf x}_{\alpha}-{\bf x}_{\beta}}{r_{\alpha \beta}},
\end{equation}
is parallel to the direction ${\bf x}_{\alpha}-{\bf x}_{\beta}$ and satisfies ${\bf f}_{\alpha \beta}=-{\bf f}_{\beta \alpha}$. 

In order to bring our approach closer to more concrete results we may use the Gaussian weighting function 
\begin{equation}
\hat{\omega}(r)=\pi^{-3/2}r^{-3}_{\omega} e^{-\frac{r^2}{r_{\omega}}},
\end{equation}
where $r_{\omega}$ is a specified radius related with the Raman probe, while for the pair potential we may utilize the Mie potential. This intermolecular pair potential between two particles at a distance $r$ is written as
\begin{equation}
V_{\alpha \beta}=\frac{n}{n-m} (\frac{n}{m})^{\frac{m}{n-m}} \varepsilon \left[ \left( \frac{\mu}{r}\right)^n-\left( \frac{\mu}{r} \right)^m \right],
\end{equation}
where $r=r_{\alpha}-r_{\beta}$, $\mu$ is the value of $r$ at $V(r)=0$, $\varepsilon$ is the well depth. Clearly, when $n=12, m=6$ it becomes the Lennard-Jones potential. The minimum of the Mie potential is located at $r_{min}=\left( \frac{n}{m} \mu^{n-m}\right)^{\frac{1}{n-m}} $. 

Substituting the bond function to the Hardy stress tensor expression we obtain 
\begin{equation}
\boldsymbol \sigma^{\text{v}}_{\omega}({\bf x}, t)=\frac{1}{2 \tau} \sum_{\alpha, \beta, \alpha \neq \beta} \int_t^{t+\tau}[- \frac{\partial V_{int}}{\partial \zeta_{\alpha \beta}} \frac{{\bf x}_{\alpha}-{\bf x}_{\beta}}{r_{\alpha \beta}} \otimes ({\bf x}_{\alpha}-{\bf x}_{\beta})] \int_{s=0}^1 \omega((1-s){\bf x}_{\alpha}+s{\bf x}_{\beta}-{\bf x}) ds dt.  
\end{equation}
For term $\frac{\partial V_{\text{int}}}{\partial \zeta_{\alpha \beta}}$ we evaluate
\begin{equation}
\frac{\partial V_{\alpha \beta}(r)}{\partial r}=\frac{n}{n-m} (\frac{n}{m})^{\frac{m}{n-m}} \varepsilon \left[ \left( \frac{\mu^n n}{r^{n+1}}\right)-\left( \frac{\mu^m m}{r^{m+1}} \right) \right].
\end{equation}
If eq. (52) is substituted to eq. (51) the stress tensor render
\begin{equation}
\boldsymbol \sigma^{\text{v}}_{\omega}({\bf x}, t)=\frac{1}{2 \tau} \sum_{\alpha, \beta, \alpha \neq \beta} \int_t^{t+\tau}[- \frac{\partial V_{int}}{\partial \zeta_{\alpha \beta}} \frac{{\bf x}_{\alpha}-{\bf x}_{\beta}}{r_{\alpha \beta}} \otimes ({\bf x}_{\alpha}-{\bf x}_{\beta})] \int_{s=0}^1 \pi^{-3/2} r_{\omega}^{-3} e^{-\frac{||(1-s){\bf x}_{\alpha}+s{\bf x}_{\beta}-{\bf x}||^2}{r_{\omega}}} ds dt.
\end{equation}

Eq. (53) expresses the Hardy stress tensor and the character of its components stem from term ${\bf x}_{\alpha}-{\bf x}_{\beta} \otimes ({\bf x}_{\alpha}-{\bf x}_{\beta})$. This term is the one which determines whether we speak about axial ($\sigma_{11}$ or $\sigma_{22}$) or shear ($\sigma_{12}$) stress components. For the case of pair potential essentially one has ${\bf z}_{\alpha \beta}={\bf x}_{\alpha}-{\bf x}_{\beta}$. Thus, for the $A_{1g}$ the findings of Section 3.2 can be used which finally render for the stress tensor for point A
\begin{eqnarray}
\boldsymbol \sigma^{\text{v}}_{\omega}({\bf x}, t)&=&\frac{1}{2 \tau} \epsilon \int_t^{t+\tau} \delta u [- \frac{\partial V_{int}}{\partial \zeta_{\alpha \beta}} \frac{10,703 {\bf z}_1 \otimes {\bf z}_1+20,975 ({\bf z}_1 \otimes {\bf z}_2+{\bf z}_2 \otimes {\bf z}_1)+15,412{\bf z}_2 \otimes {\bf z}_2}{6,142}] \nonumber\\
&&\pi^{-3/2} r_{\omega}^{-3}  dt.
\end{eqnarray}
To proceed one should choose values for the two time scales $t, \tau$, for the specified radius $r_{\omega}$ as well as for $m, n$ to specify the pair potential. Also, for the term $\delta u$ a plane wave assumption of the form $\delta u=e^{-i \omega \tau}$ can be done in similar trend with \cite{Sahataetal1988}. After these choices are made the last formula for $\boldsymbol \sigma^{\text{v}}_{\omega}({\bf x}, t)$ can be straightforwardly evaluated rendering the stress tensor. Clearly, every kind of stress components, axial as well as shear, result as an outcome of the $A_{1g}$ vibration mode. So, the main outcome is again that when $A_{1g}$ mode is considered all stress tensor components should be taken into account. Essentially, what is achieved in this Section compared to the considerations of Section 3 is that ensemble average is substituted by time average, thus calculations can be done, since the phase space of Section 3 is in general not known. 

In a similar manner, for the $E_{2g}$ vibration mode, considerations of Section 3.3 can be used to find for the Hardy stress tensor corresponding to this mode of vibration for point A the formula
\begin{eqnarray}
\boldsymbol \sigma^{\text{v}}_{\omega}({\bf x}, t)&=&\frac{1}{2 \tau}  \epsilon \int_t^{t+\tau} \delta u [- \frac{\partial V_{int}}{\partial \zeta_{\alpha \beta}} \frac{17,804 {\bf z}_1 \otimes {\bf z}_1+15,046 ({\bf z}_1 \otimes {\bf z}_2+{\bf z}_2 \otimes {\bf z}_1)}{27,744}] \nonumber\\
&&\pi^{-3/2} r_{\omega}^{-3}  dt.
\end{eqnarray}
Calculations can be further pursued by making choices for $t, \tau$ as for eq. (54). Similarly with Section 3.3 we infer that when $E_{2g}$ mode is activated axial stress components of the form $\sigma_{11}$ as well as shear stress components of the form of $\sigma_{12}$ are created. Axial stresses stem from term $17,804 {\bf z}_1 \otimes {\bf z}_1 $ while shear stresses stem from terms $15,046 ({\bf z}_1 \otimes {\bf z}_2+{\bf z}_2 \otimes {\bf z}_1)$. 

\section{Applied external field}

When an external stress or tension field applies to the graphene ring there are some important questions that naturally arise. Firstly, one has to treat the way the external field applies compared to the inclination of the material and how this is distributed within the specimen. We here assume that the distribution of the applied external field is done homogeneously within the material. For the inclination of the applied external field we study homogeneous tension along the armchair and the zig-zag direction and we then generalize to an arbitrary direction. 

Two question are also crucial: i) what is the degree of the perturbation compared to the applied external field, and ii) does the perturbation changes due to the external field? For the first question is seems natural to assume that the degree of the perturbation is very small compared to the applied external field. Of course, to safely support such an argument one has to measure the stress components from the perturbations and the same quantities that arise from the applied external field and compare their values. Unfortunately, this cannot be done. Nevertheless, in classical problems of stability (\cite{Sfyris2011}), the perturbation $e^{i(k {\bf n} \cdot {\bf x}-\omega t)}$ is of a small order. In standard Raman works such as \cite{Sahataetal1988} stresses range from 0 to 430 MPa so it seems safely to assume that the perturbation is of smaller order compared to the external field. 

Concerning the second question, it seems that the order of the perturbation remains the same when the external field applies. Of course the frequency of the perturbation changes, but this change is very small. This can be justified from Figure 2 of \cite{Sahataetal1988} where it is seen that for applied external stress field ranging from 0 to 430 MPa, frequency changes less that 5cm$^{-1}$. Thus, it seems natural to assume that the perturbation changes insignificantly due to the external loading. From the mathematical point of view this means that $\epsilon$ in front of the perturbation is of the same order as in the previous calculations. In Section 5.1 we study tension along the armchair direction while in Section 5.2 tension is along the zig-zag. In Section 5.3 we generalize the tension to be at an arbitrary direction.  

\subsection{Tension along the armchair direction}
By assuming that the applied external tension is homogeneously distributed along the armchair direction, namely of the form of Figure 4,
\begin{figure}[!htb]
\centering
\includegraphics[width=80mm]{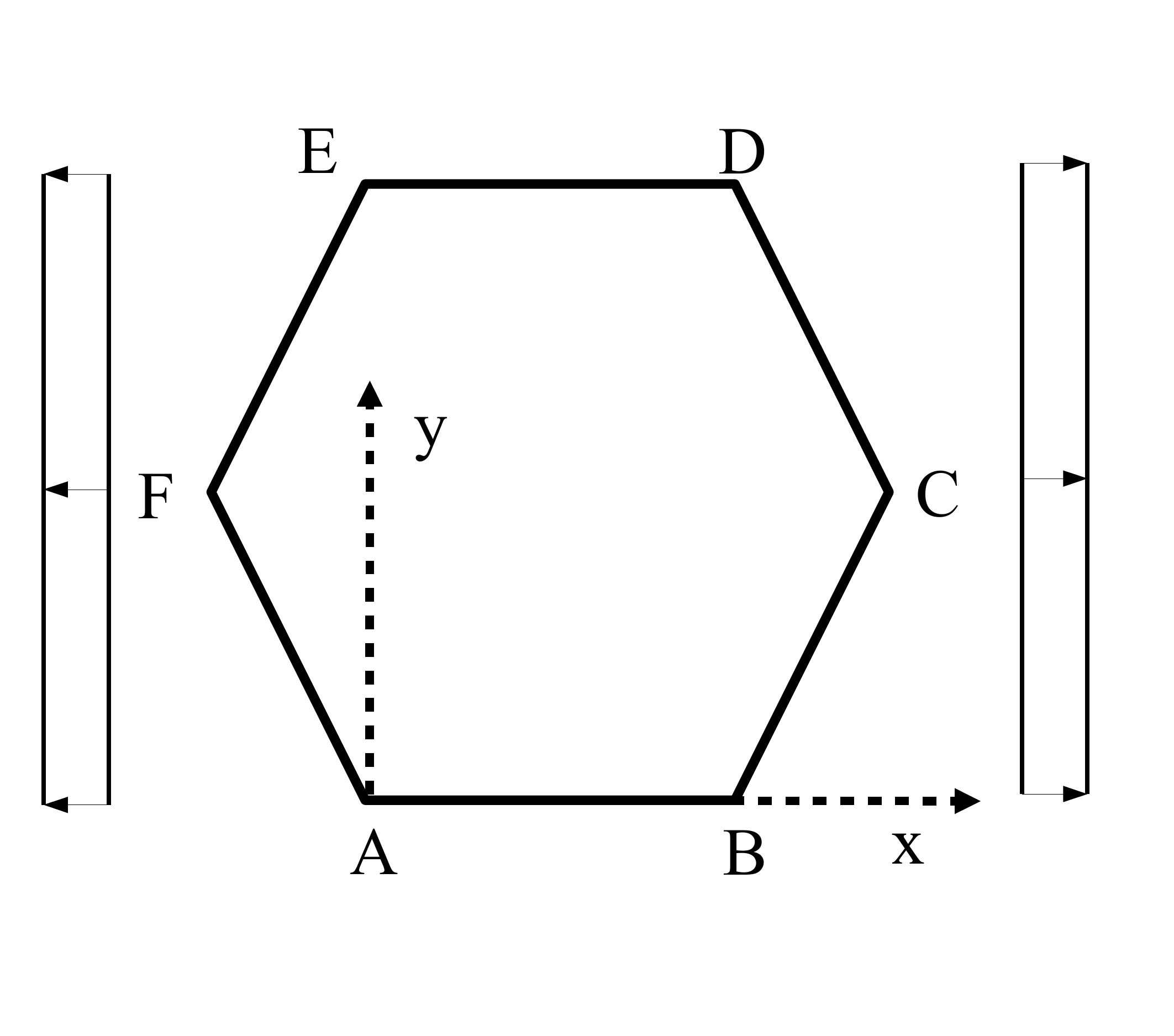}
\caption{Applied tension along the armchair direction.  }
\end{figure}
the hexagonal ring elongates along the x-direction with bond lengths AB, AC, AD being affected only. Bonds AE, AF do not change either in shape or in length. So, what is new in the tensor product ${\bf z}^* \otimes {\bf z}^*$ is that new terms should be added in the ${\bf z}_{AB}^*$, ${\bf z}_{AC}^*$ and ${\bf z}_{AD}^*$ terms. So, with the applied strain ${\bf z}^{\text{appl}}$ the bonds that change take the following expression for the $A_{1g}$
\begin{eqnarray}
{\bf z}_{AB}&=&(1,42  +z^{\text{appl}}) {\bf z}_1,  \\
{\bf z}_{AC}&=&(2,13+z^{\text{appl}}) {\bf z}_1+1,229 {\bf z}_2,  \\
{\bf z}_{AD}&=&(2,13+z^{\text{appl}}) {\bf z}_1+2,258 {\bf z}_2.
\end{eqnarray}

When the perturbations are introduced as well from Section 3.2 we additionally run into
\begin{eqnarray}
{\bf z}_{AB}^*&=&(1,42+z^{\text{appl}}+\epsilon \delta u) {\bf z}_1+0,866 \epsilon \delta u{\bf z}_2, \\
{\bf z}_{AC}^*&=&(2,13+z^{\text{appl}}+1,5\epsilon \delta u) {\bf z}_1+(1,229+0,866 \epsilon \delta u) {\bf z}_2, \\
{\bf z}_{AD}^*&=&(2,13+z^{\text{appl}}+\epsilon \delta u) {\bf z}_1+(2,258+0,732 \epsilon \delta u) {\bf z}_2.
\end{eqnarray}
Terms ${\bf z}_{AF}^*$, ${\bf z}_{AE}^*$ remain the same as that of Section 3.2, since the corresponding bonds do not alter. By forming the product ${\bf z}^* \otimes {\bf z}^*$ after some lengthy but straightforward calculations we obtain 
\begin{eqnarray}
&&{\bf z}^* \otimes {\bf z}^*={\bf z}_{AB}^* \otimes {\bf z}_{AB}^*+{\bf z}_{AC}^* \otimes {\bf z}_{AC}^*+{\bf z}_{AD}^* \otimes {\bf z}_{AD}^*+{\bf z}_{AE}^* \otimes {\bf z}_{AE}^*+{\bf z}_{AF}^* \otimes {\bf z}_{AF}^* \nonumber\\  
&&=\{ 8,063+11,36 z^{\text{appl}}+ \epsilon \delta u(14,293 +4 z^{\text{appl}}) +5,5 (\epsilon \delta u)^2 + 3 z^{\text{appl}}  \} {\bf z}_1 \otimes {\bf z}_1 \nonumber\\
&&\{ 5,915+\epsilon \delta u(9,109 +2,464  z^{\text{appl}}) +3,847 z^{\text{appl}}+2,662 (\epsilon \delta u)^2 \} ({\bf z}_1 \otimes {\bf z}_2+{\bf z}_2 \otimes {\bf z}_1) \nonumber\\
&&\{ 4,698+5,433 \epsilon \delta u +1,358 (\epsilon \delta u)^2  \} {\bf z}_2 \otimes {\bf z}_2 .
\end{eqnarray}

So, the stress at point A of the hexagonal ring evaluates
\begin{eqnarray}
\boldsymbol \sigma^{\text{v}}({\bf x}, t)&=&\boldsymbol \sigma^{\text{v}}_{\text{int}}+\int_{\mathcal R^3} \frac{1}{||{\bf z}^* \otimes {\bf z}^*||} \nonumber\\
&& (\{ 8,063+11,36 z^{\text{appl}}+ \epsilon \delta u (14,293+4 z^{\text{appl}}) +5,5 (\epsilon \delta u)^2 + 3 z^{\text{appl}}  \} {\bf z}_1 \otimes {\bf z}_1+ \nonumber\\
&&\{ 5,915+ \epsilon \delta u(9,109 +2,464 z^{\text{appl}}) +3,847 z^{\text{appl}}+2,662 (\epsilon \delta u)^2 \} ({\bf z}_1 \otimes {\bf z}_2+{\bf z}_2 \otimes {\bf z}_1)+ \nonumber\\
&&\{ 4,698+5,433 \epsilon \delta u +1,358 (\epsilon \delta u)^2  \} {\bf z}_2 \otimes {\bf z}_2) \\
&&<\frac{\partial V_{\text{int}}}{\partial r_{\alpha \beta}}W|{\bf x}_{\alpha}={\bf x}, {\bf x}_{\beta}={\bf x}-{\bf z}> d{\bf x}. \nonumber
\end{eqnarray}
By assuming that the internal stresses of Section 3.1, $\boldsymbol \sigma^{\text{v}}_{\text{int}}$, are negligible and also that the perturbations are of smaller order compared to the applied strain, we see that the axial $\sigma_{11}$ stress component (the one related with ${\bf z}_1 \otimes {\bf z}_1$) has term $(z^{\text{appl}})^2$ in its expression: this is the highest order term. Term $z^{\text{appl}}$  appears also in the shear $\sigma_{12}$ component (the one related with ${\bf z}_1 \otimes {\bf z}_2$) but not squared, while the other axial component, i.e. $\sigma_{22}$ (the one related with ${\bf z}_2 \otimes {\bf z}_2$) does not have $z^{\text{appl}}$ in its expression. It is obvious that $\sigma_{11}$ grows faster than $\sigma_{12}$ as $z^{\text{appl}}$ becomes greater in the same way that the function $x^2$ grows faster than function $x$, when $x \geq 1$.  Namely, for our case, if $z^{\text{appl}}$ is of the order of some decades of Angstrom then all terms are important (even $\sigma_{22}$), since they are all of the same order. If we apply tension more than some Angstrom then the axial component $\sigma_{11}$ dominates and it always is at the square scale of $z^{\text{appl}}$. So, all in all, for axial tension along the armchair direction and for case when tension is more than some Angstrom, the axial $\sigma_{11}$ is the dominant one, while the shear stress $\sigma_{12}$ component is of lower order. A full analysis should take both quantities into account, but in a first approximation only the dominant $\sigma_{11}$ should be used as is correctly done in \cite{Sahataetal1988}. 

In some experiments it is the applied stress which is controlled, so it may be useful to change $z^{\text{appl}}$ to $\sigma^{\text{appl}}$. One way to do this is to use Hooke's law which for our case for bond AB reads
\begin{equation}
{\boldsymbol \sigma}^{\text{appl}}=E \frac{{\bf z}_{AB}^{\text{final}}-{\bf z}_{AB}^{\text{initial}}}{{\bf z}_{AB}^{\text{initial}}}= E \frac{{\bf z}_{AB}^{\text{appl}}}{{\bf z}_{AB}^{\text{initial}}}
\end{equation}
so by setting $|{\bf z}_{AB}^{\text{initial}}|=1$ after a suitable rescaling, one may use
\begin{equation}
{\bf z}_{AB}^{\text{appl}}=\frac{{\boldsymbol \sigma^{\text{appl}}}}{E},
\end{equation}
$E$ being graphene's Young modulus.
After such an interchange one can reiterate the discussion for the dominant term in terms of $\sigma^{\text{appl}}$ rather than $z^{\text{appl}}$ with exactly the same outcome. We note here that use of Hooke's law as that of eq. (65) tacitly assumes that we are in the linear regime, a fact that restricts the applicability of this discussion and should be duly taken into account when used. 

For the $E_{2g}$ mode taking into account calculations of Section 3.3 for the perturbations and the fact that bonds AB, AC, AD change only, we have  
\begin{eqnarray}
{\bf z}^*_{AB}&=&(1,42+z^{\text{appl}} + 2 \epsilon \delta u) {\bf z}_1, \\
{\bf z}^*_{AC}&=&(2,13+z^{\text{appl}} + 2 \epsilon \delta u) {\bf z}_1+1,229 {\bf z}_2, \\
{\bf z}^*_{AD}&=&(2,13+z^{\text{appl}} + 2 \epsilon \delta u) {\bf z}_1+2,258 {\bf z}_2, \\
{\bf z}^*_{AE}&=&( 2 \epsilon \delta u) {\bf z}_1+2,258 {\bf z}_2, \\
{\bf z}^*_{AF}&=&(-1,229 + 2 \epsilon \delta u) {\bf z}_1+1,229 {\bf z}_2. 
\end{eqnarray}
So, after some lengthy but straightforward calculations we obtain for the tensor product ${\bf z}^* \otimes {\bf z}^*$ that
\begin{eqnarray}
{\bf z}^* \otimes {\bf z}^*&=& (12,6006+11,36 z^{\text{appl}}+\epsilon \delta u(6,45+10,84 z^{\text{appl}})+20 (\epsilon \delta u)^2+5 (z^{\text{appl}})^2) {\bf z}_1 \otimes {\bf z}_1 \nonumber\\
&&+(5,9163+3,487z^{\text{appl}}+13,948 \epsilon \delta u)({\bf z}_1 \otimes {\bf z}_2+{\bf z}_2 \otimes {\bf z}_1) \\
&&+13,2168 {\bf z}_2 \otimes {\bf z}_2. \nonumber
\end{eqnarray}
Thus, the stress tensor for this case reads
\begin{eqnarray}
\boldsymbol \sigma^{\text{v}}({\bf x}, t)&=&\boldsymbol \sigma^{\text{v}}_{\text{int}}+ \int_{\mathcal R^3} \frac{1}{||{\bf z}^* \otimes {\bf z}^*||} \nonumber\\
&& (\{ 12,6006+11,36 z^{\text{appl}}+\epsilon \delta u(6,45+10,84 z^{\text{appl}})+20 (\epsilon \delta u)^2+5 (z^{\text{appl}})^2  \} {\bf z}_1 \otimes {\bf z}_1+ \nonumber\\
&&\{ 5,9163+3,487z^{\text{appl}}+13,948 \epsilon \delta u\} ({\bf z}_1 \otimes {\bf z}_2+{\bf z}_2 \otimes {\bf z}_1)+ \nonumber\\
&&\{ 13,2168  \} {\bf z}_2 \otimes {\bf z}_2) \\
&&<\frac{\partial V_{\text{int}}}{\partial r_{\alpha \beta}}W|{\bf x}_{\alpha}={\bf x}, {\bf x}_{\beta}={\bf x}-{\bf z}> d{\bf x}. \nonumber
\end{eqnarray}
Inspecting terms of the above equation we see that it is the axial $\sigma_{11}$ component that contain term $(z^{\text{appl}})^2$, while the shear component $\sigma_{12}$ contain term $z^{\text{appl}}$. By a similar analysis as before we therefore infer that the axial component $\sigma_{11}$ dominates over $\sigma_{12}, \sigma_{22}$ stress components for applied strain greater than that of some decades of Angstrom. 

\subsection{Tension along the zig-zag direction}

Tension along the zig-zag direction is assumed to be applied homogeneously as seen in Figure 5.
\begin{figure}[!htb]
\centering
\includegraphics[width=80mm]{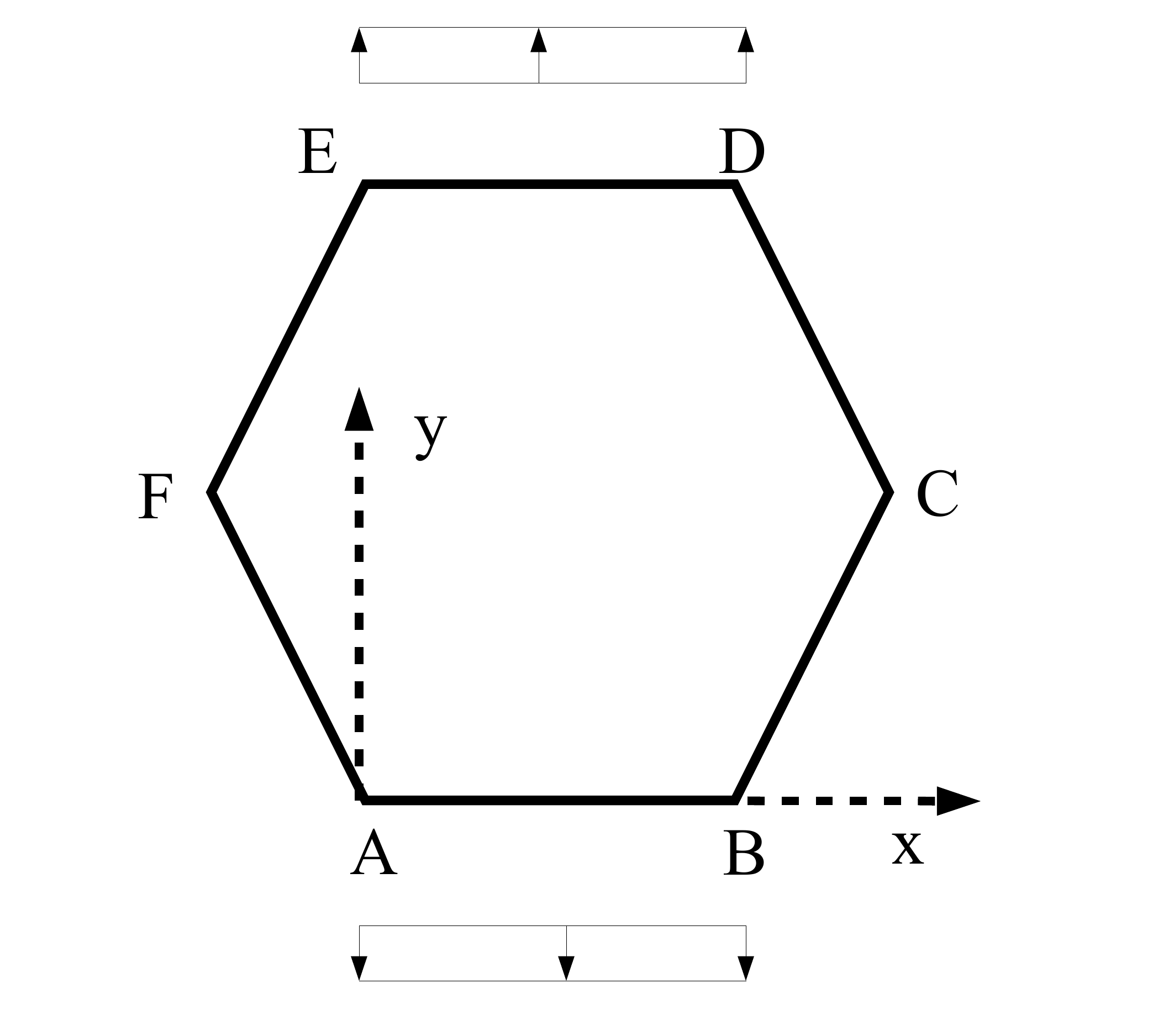}
\caption{Applied tension along the zig-zag direction.  }
\end{figure}
For the $A_{1g}$ mode taking into account that only bond AB remain as is we find using as well the perturbation part of Section 3.2 that
\begin{eqnarray}
{\bf z}^*_{AB}&=&(1,42+\epsilon \delta u) {\bf z}_1+0,866\epsilon \delta u {\bf z}_1, \\
{\bf z}^*_{AC}&=&(2,13-\nu z^{\text{appl}} + 1,5 \epsilon \delta u) {\bf z}_1+(1,229+z^{\text{appl}}+0,866\epsilon \delta u) {\bf z}_2, \\
{\bf z}^*_{AD}&=&(2,13+ \epsilon \delta u) {\bf z}_1+(2,258+0,732 \epsilon \delta u+2 z^{\text{appl}}) {\bf z}_2, \\
{\bf z}^*_{AE}&=&( \epsilon \delta u) {\bf z}_1+(2,258+0,732\epsilon \delta u+2z^{\text{appl}}) {\bf z}_2, \\
{\bf z}^*_{AF}&=&(-1,229 +\nu z^{\text{appl}}+ 1,5 \epsilon \delta u) {\bf z}_1+(1,229+z^{\text{appl}}+0,866 \epsilon \delta u) {\bf z}_2. 
\end{eqnarray}
For the tensor product ${\bf z}^* \otimes {\bf z}^*$ in this case we obtain
\begin{eqnarray}
{\bf z}^* \otimes {\bf z}^*&=& (12,6002+\epsilon \delta u+7,5 (\epsilon \delta u)^2-6,779 \nu z^{\text{appl}}+2 \nu^2 z^{2\text{appl}}) {\bf z}_1 \otimes {\bf z}_1 \nonumber\\
&&+(5,9156+\epsilon \delta u(11,797+6,484 z^{\text{appl}})+4,843 (\epsilon \delta u)^2-1,229 \nu z^{\text{appl}}+4,262 z^{\text{appl}}) \nonumber\\
&&({\bf z}_1 \otimes {\bf z}_2+{\bf z}_2 \otimes {\bf z}_1) \\
&&+(11,706+\epsilon \delta u(20,864+9,32 z^{\text{appl}})+3,31 (\epsilon \delta u)^2+23,0412 z^{\text{appl}}+10 z^{2\text{appl}}) {\bf z}_2 \otimes {\bf z}_2. \nonumber
\end{eqnarray}

It is interesting here to explain how the Poisson ratio, $\nu$, of graphene appear in the above relations. We can immediately see that from Figure 6. 
\begin{figure}[!htb]
\centering
\includegraphics[width=100mm]{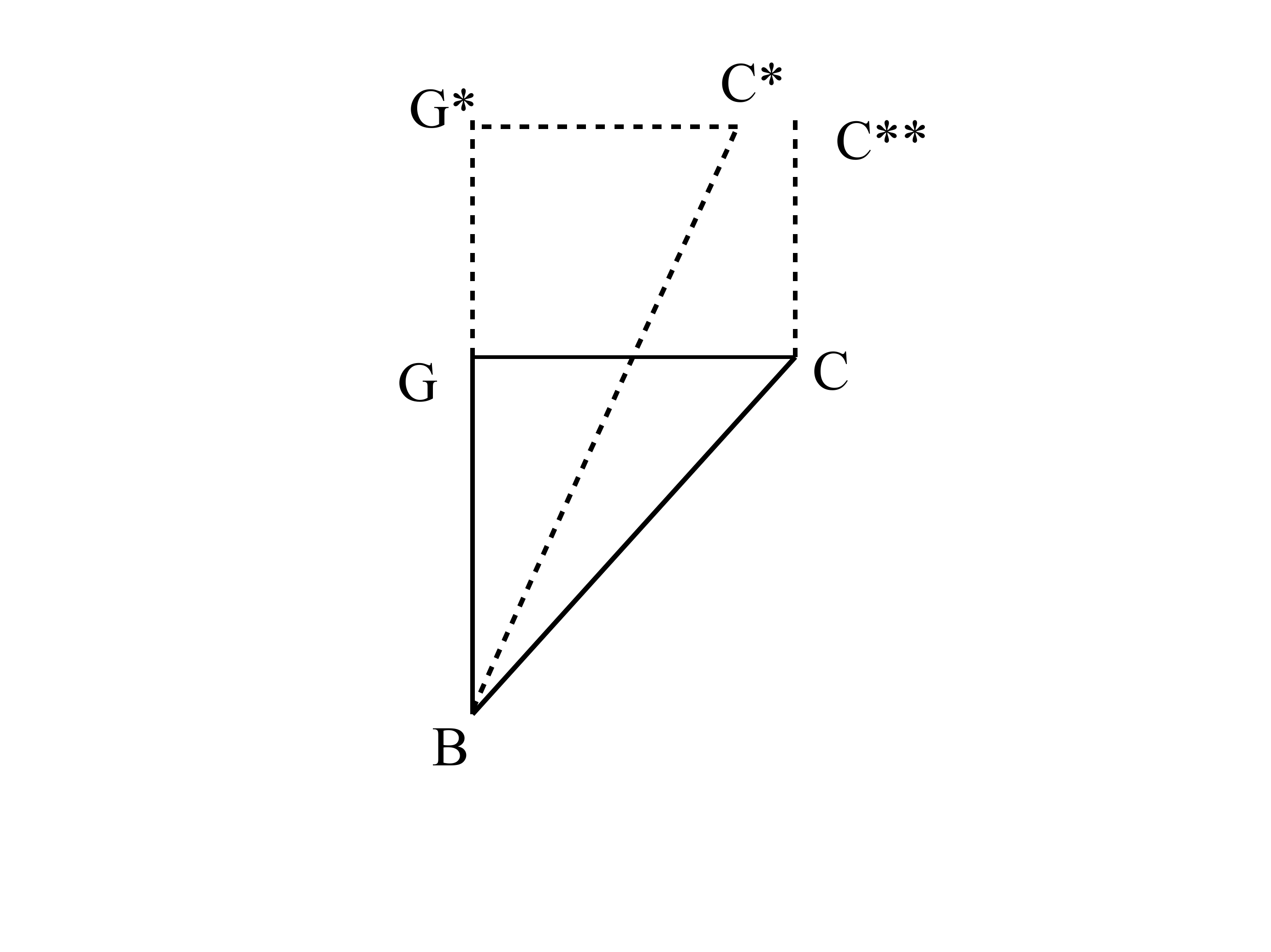}
\caption{The bond BC before and after the application of the tensile field.  }
\end{figure}
There the solid line represents the AC bond before tension applies. Dashed line represents the deformed case. The $GG^*$ is $z^{\text{appl}}$, while $C^*C^{**}$ is how much the ring shrinks in the transverse direction when $z^{\text{appl}}$ applies which from linear elasticity and the standard definition of the Poisson ratio is $-\nu z^{\text{appl}}$. We also note that use of the $\nu$ tacitly assumes that the material behaves in a linear elastic manner since Poisson ratio is a constant used in the linear elastic regime only (\cite{Scott2007}). 

The stress tensor in this case becomes
\begin{eqnarray}
\boldsymbol \sigma^{\text{v}}({\bf x}, t)&=&\boldsymbol \sigma^{\text{v}}_{\text{int}}+ \int_{\mathcal R^3} \frac{1}{||{\bf z}^* \otimes {\bf z}^*||}\nonumber\\
&& (\{ 12,6002+\epsilon \delta u+7,5 (\epsilon \delta u)^2-6,779 \nu z^{\text{appl}}+2 (\nu z^{\text{appl}})^2  \} {\bf z}_1 \otimes {\bf z}_1+ \nonumber\\
&&\{ 5,9156+\epsilon \delta u(11,797+6,484 z^{\text{appl}})+4,843 (\epsilon \delta u)^2-1,229 \nu z^{\text{appl}}+4,262 z^{\text{appl}}\} \nonumber\\
&& ({\bf z}_1 \otimes {\bf z}_2+{\bf z}_2 \otimes {\bf z}_1)+ \nonumber\\
&&\{ 11,706+\epsilon \delta u(20,864+9,32 z^{\text{appl}})+3,31 (\epsilon \delta u)^2+23,0412 z^{\text{appl}}+10 (z^{\text{appl}})^2 \} {\bf z}_2 \otimes {\bf z}_2) \nonumber\\
&&<\frac{\partial V_{\text{int}}}{\partial r_{\alpha \beta}}W|{\bf x}_{\alpha}={\bf x}, {\bf x}_{\beta}={\bf x}-{\bf z}> d{\bf x}.
\end{eqnarray}
The axial $\sigma_{11}$ component contain term $(\nu z^{\text{appl}})^2$, $\sigma_{12}$ shear stress component contain term  $z^{\text{appl}}$ and the axial $\sigma_{22}$ stress component contain term $(z^{\text{appl}})^2$. Granted that for graphene in the linear regime the Poisson ratio ranges from 0.14-0.42 (\cite{SfyrisetalJAP2015}) with a mean value being approximately 0.3, we infer that component $\sigma_{22}$ dominates, with component $\sigma_{11}$ being two orders of magnitude smaller and $\sigma_{12}$ being of order of $z^{\text{appl}}$.

For the $E_{2g}$ mode working in a similar line of thought we have
\begin{eqnarray}
{\bf z}^*_{AB}&=&(1,42+2\epsilon \delta u) {\bf z}_1, \\
{\bf z}^*_{AC}&=&(2,13-\nu z^{\text{appl}} + 2 \epsilon \delta u) {\bf z}_1+(1,229+z^{\text{appl}}) {\bf z}_2, \\
{\bf z}^*_{AD}&=&(2,13+2 \epsilon \delta u) {\bf z}_1+(2,258+2 z^{\text{appl}}) {\bf z}_2, \\
{\bf z}^*_{AE}&=&(2 \epsilon \delta u) {\bf z}_1+(2,258+2z^{\text{appl}}) {\bf z}_2, \\
{\bf z}^*_{AF}&=&(-1,229 +\nu z^{\text{appl}}+ 2 \epsilon \delta u) {\bf z}_1+(1,229+z^{\text{appl}}) {\bf z}_2. 
\end{eqnarray}
So, for the tensor product we obtain
\begin{eqnarray}
{\bf z}^* \otimes {\bf z}^*&=& (15,087+\epsilon \delta u (25,218+8 \nu z^{\text{appl}})+20 (\epsilon \delta u)^2-6,718 \nu z^{\text{appl}}+2 (\nu z^{\text{appl}})^2) {\bf z}_1 \otimes {\bf z}_1 \nonumber\\
&&+(8,6441+\epsilon \delta u (9,432+12 z^{\text{appl}})+5,161 z^{\text{appl}}))({\bf z}_1 \otimes {\bf z}_2+{\bf z}_2 \otimes {\bf z}_1) \\
&&+(13,2168+10 (z^{\text{appl}})^2+22,98 z^{\text{appl}}) {\bf z}_2 \otimes {\bf z}_2. \nonumber
\end{eqnarray}

The stress tensor for this case then reads
\begin{eqnarray}
\boldsymbol \sigma^{\text{v}}({\bf x}, t)&=&\boldsymbol \sigma^{\text{v}}_{\text{int}}+ \int_{\mathcal R^3} \frac{1}{||{\bf z}^* \otimes {\bf z}^*||}\nonumber\\
&& (\{ 15,087+\epsilon \delta u (25,218+8 \nu z^{\text{appl}})+20 (\epsilon \delta u)^2-6,718 \nu z^{\text{appl}}+2 (\nu z^{\text{appl}})^2  \} {\bf z}_1 \otimes {\bf z}_1+ \nonumber\\
&&\{ 8,6441+\epsilon \delta u (9,432+12 z^{\text{appl}})+5,161 z^{\text{appl}})\} ({\bf z}_1 \otimes {\bf z}_2+{\bf z}_2 \otimes {\bf z}_1)+ \nonumber\\
&&\{ 13,2168+10 (z^{\text{appl}})^2+22,98 z^{\text{appl}}  \} {\bf z}_2 \otimes {\bf z}_2) \\
&&<\frac{\partial V_{\text{int}}}{\partial r_{\alpha \beta}}W|{\bf x}_{\alpha}={\bf x}, {\bf x}_{\beta}={\bf x}-{\bf z}> d{\bf x},
\end{eqnarray}
The axial $\sigma_{22}$ component of the stress is the dominant one since it contain term $(z^{\text{appl}})^2$. The shear component $\sigma_{12}$ contains term $z^{\text{appl}}$, so it is of the order of $z^{\text{appl}}$, thus of a lower order from $\sigma_{22}$. The axial component $\sigma_{11}$ contains term $(\nu z^{\text{appl}})^2$, and granted that $\nu=0.14-0.42$ with a mean value of 0.3 approximately, $\sigma_{11}$ is two orders of magnitude smaller than the other axial component. 

\subsection{Tension along an arbitrary direction}

To generalize our analysis to more realistic situations, we examine the case where the tensile loading is applied homogeneously in a direction which makes a $\theta$ angle with respect to the armchair direction (see Figure 7). 
\begin{figure}[!htb]
\centering
\includegraphics[width=80mm]{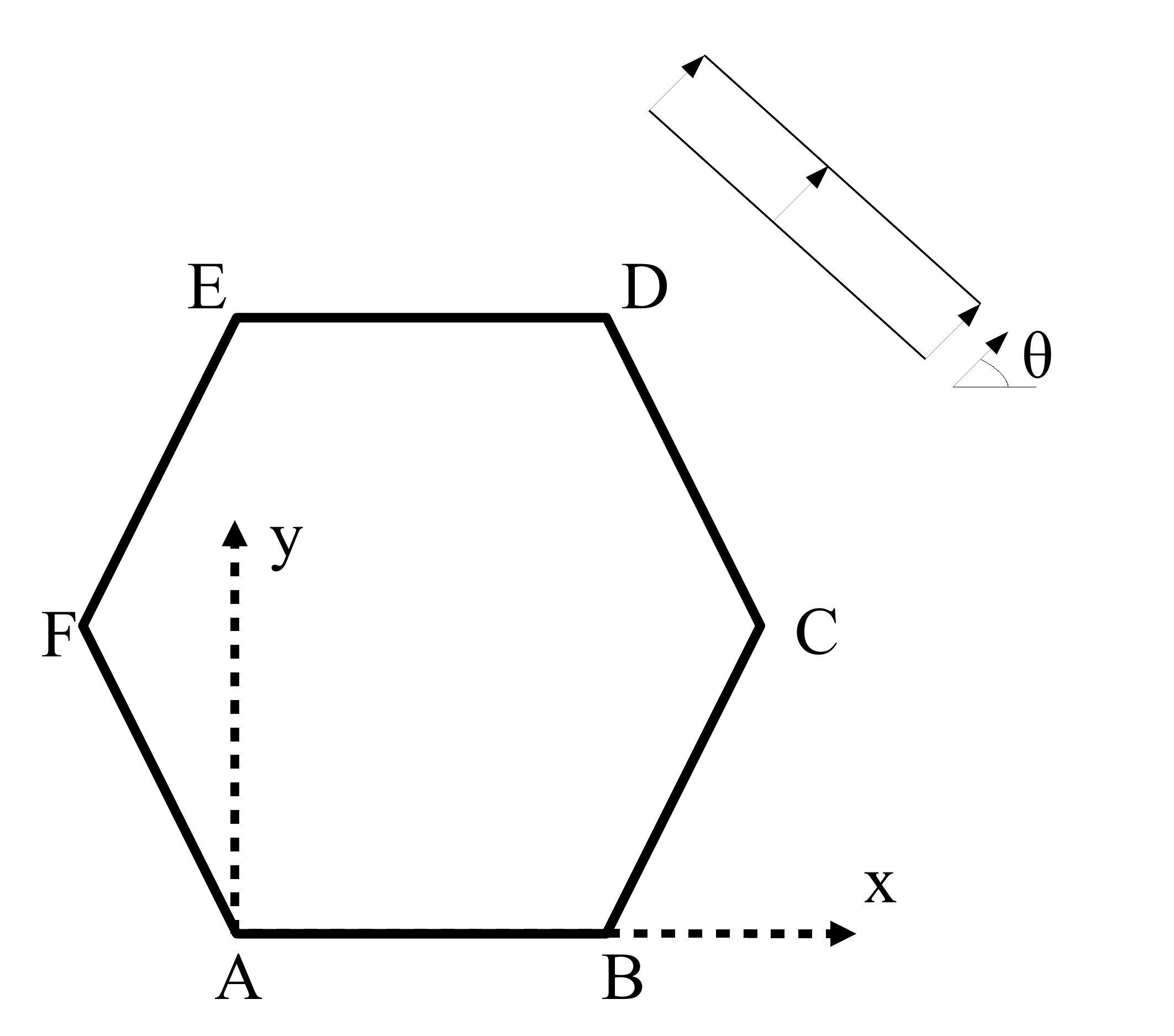}
\caption{ Applied tension at an arbitrary direction which makes a $\theta$ angle with the armchair direction. }
\end{figure}
For the triangle of the applied tension $z^{\text{appl}}$ and $\theta$ we can see Figure 8. 
\begin{figure}[!htb]
\centering
\includegraphics[width=120mm]{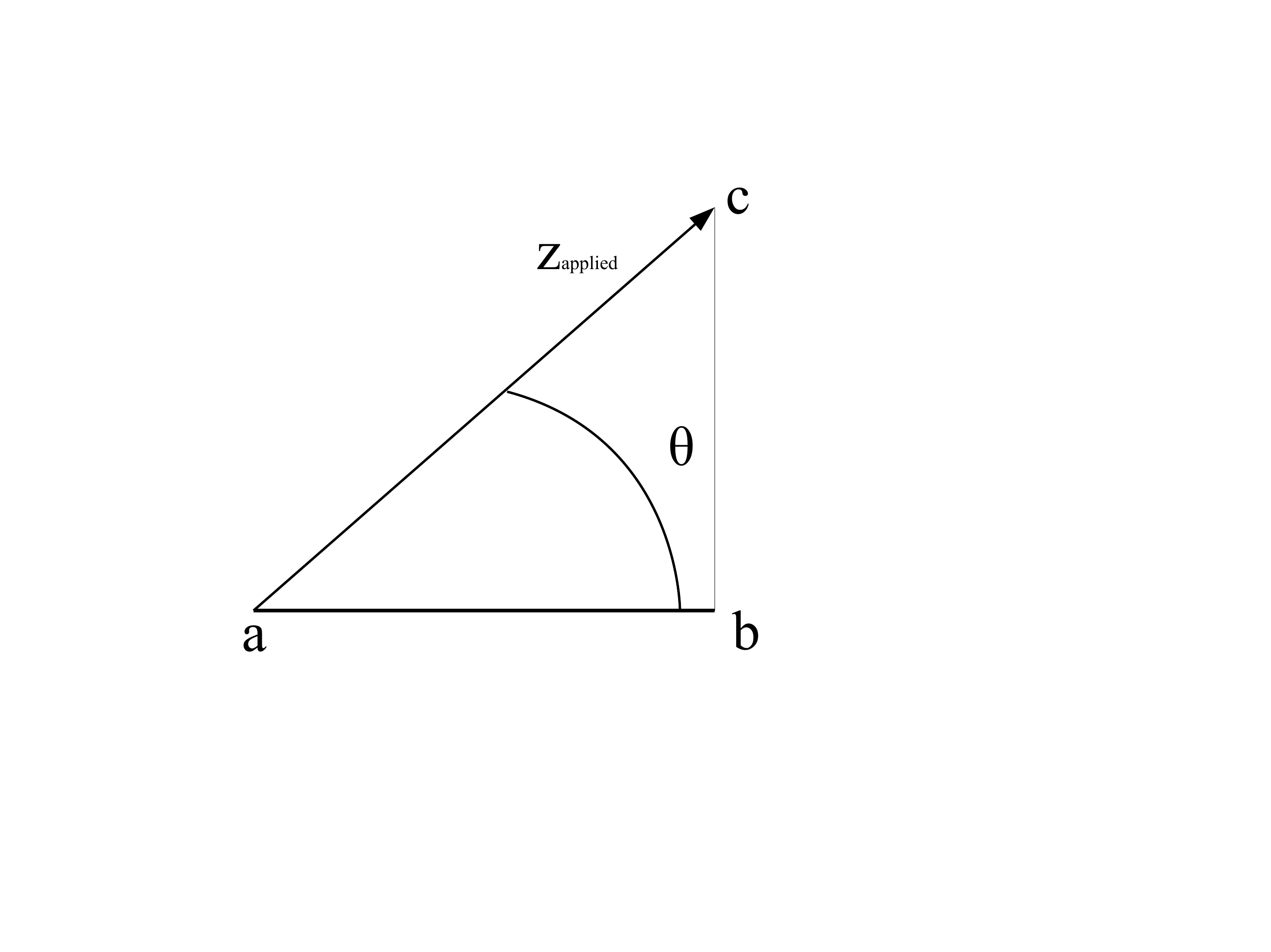}
\caption{ From the triangle of the applied loading we see that $ab=z^{\text{appl}} \text{cos} \theta$, $bc=z^{\text{appl}} \text{sin} \theta$. }
\end{figure}
There $ab=z^{\text{appl}} \text{cos} \theta$, $bc=z^{\text{appl}} \text{sin} \theta$.
\begin{figure}[!htb]
\centering
\includegraphics[width=80mm]{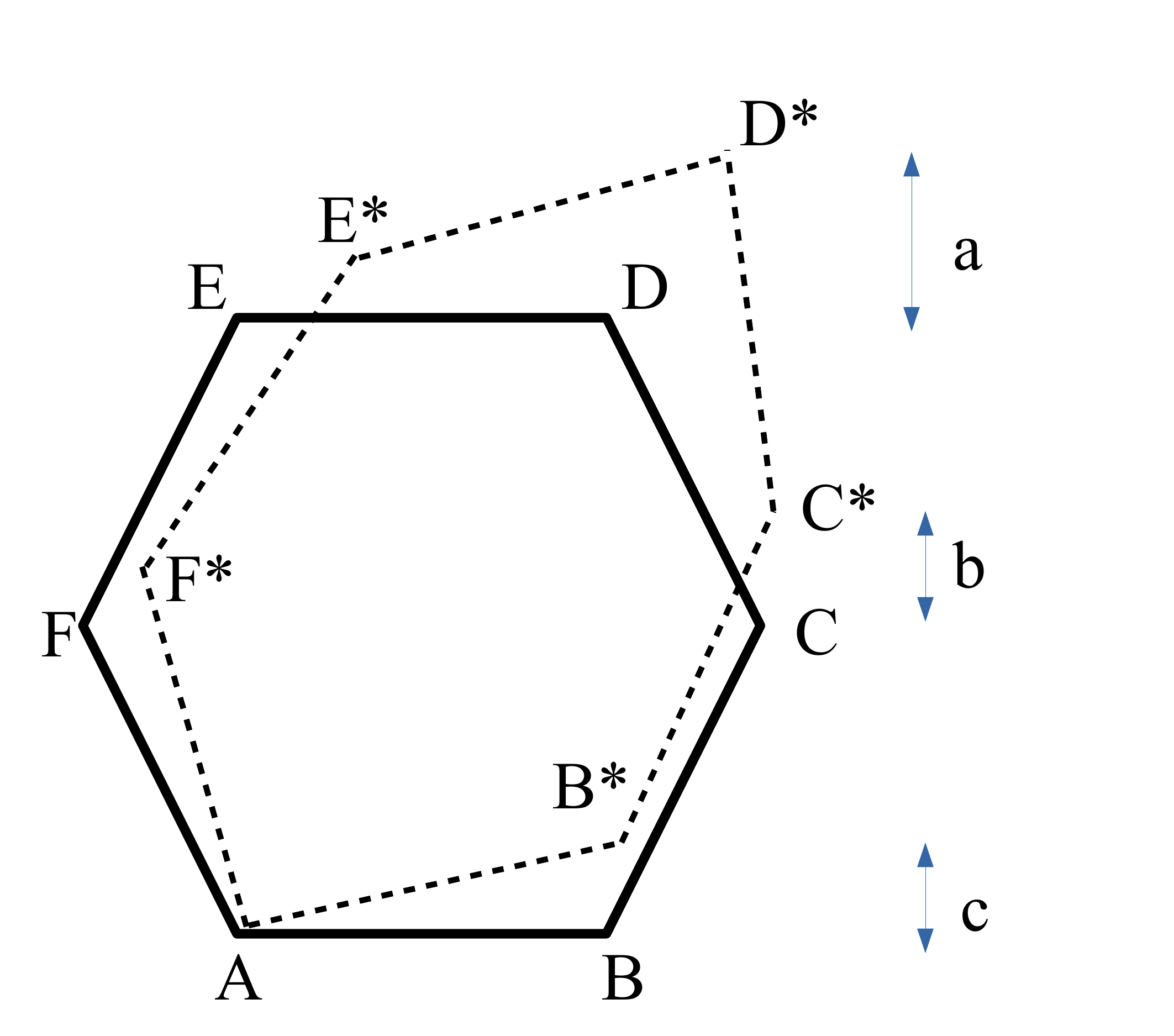}
\caption{ Distances a, b, c are different and are introduced mathematically into our analysis through terms  $c_{AB}, ..., c_{AF}, d_{AB}, ..., d_{AF}$ (see text).  }
\end{figure}
It is important to notice here that every bond experiences different deformation as an application of the tensile loading. To introduce this mathematically into our analysis, we use constants $c_{AB}, ..., c_{AF}, d_{AB}, ..., d_{AF}$ to signify the differences of lengths $a, b, c$ in Figure 9. Constants $c_{AB}, ..., c_{AF}$ measure the fraction of the deformation in the x-axis (i.e. $z^{\text{appl}} \text{cos} \theta$) that bonds AB, ..., AF, respectively, experience. Constants $d_{AB}, ..., d_{AF}$ measure the fraction of the deformation in the y-axis (i.e. $z^{\text{appl}} \text{sin} \theta$) that bonds AB, ..., AF, respectively, experience. In simple words, for each bond we have two quantities e.g. $c_{AB}, d_{AB}$ that describe changes of bond AB in the x and y-axis, respectively, due to the applied tension. 

In this case we have for the $A_{1g}$ mode 
\begin{eqnarray}
{\bf z}^*_{AB}&=&(1,42+\epsilon \delta u-c_{AB} z^{\text{appl}} \text{cos} \theta ) {\bf z}_1+(0,866\epsilon \delta u+d_{AB} z^{\text{appl}} \text{sin} \theta) {\bf z}_1, \\
{\bf z}^*_{AC}&=&(2,13-c_{AC} z^{\text{appl}} \text{cos} \theta  + 1,5 \epsilon \delta u) {\bf z}_1+(1,229+0,866\epsilon \delta u+d_{AC} z^{\text{appl}} \text{sin} \theta) {\bf z}_2, \\
{\bf z}^*_{AD}&=&(2,13+ \epsilon \delta u+c_{AD} z^{\text{appl}} \text{cos} \theta) {\bf z}_1+(2,258+0,732 \epsilon \delta u+d_{AD} z^{\text{appl}} \text{sin} \theta) {\bf z}_2, \\
{\bf z}^*_{AE}&=&( \epsilon \delta u-c_{AE} z^{\text{appl}} \text{cos} \theta) {\bf z}_1+(2,258+0,732\epsilon \delta u+d_{AE} z^{\text{appl}} \text{sin} \theta) {\bf z}_2, \\
{\bf z}^*_{AF}&=&(-1,229 -c_{AF} z^{\text{appl}} \text{cos} \theta+ 1,5 \epsilon \delta u) {\bf z}_1+(1,229+0,866 \epsilon \delta u+d_{AF} z^{\text{appl}} \text{sin} \theta) {\bf z}_2. 
\end{eqnarray}
The tensor product then reads
\begin{equation}
{\bf z}^* \otimes {\bf z}^*= A {\bf z}_1 \otimes {\bf z}_1 +B({\bf z}_1 \otimes {\bf z}_2+{\bf z}_2 \otimes {\bf z}_1) +C {\bf z}_2 \otimes {\bf z}_2, 
\end{equation}
where
\begin{eqnarray}
A&=& \{ 12,599+\epsilon \delta u(17,177+z^{\text{appl}} \text{cos} \theta (-2c_{AB}-3c_{AC}-2c_{AE}+2c_{AD}+3c_{AF})) \nonumber\\
&&+7,5 (\epsilon \delta u)^2  + z^{\text{appl}} \text{cos} \theta (-2,84 c_{AB}-4,26c_{AC}+4,26c_{AD}+2,48c_{AF}) \nonumber\\
&&+(z^{\text{appl}} \text{cos} \theta)^2 (c^2_{AB}+c^2_{AC}+c^2_{AD}+c^2_{AE}+c^2_{AE}) \}, 
\end{eqnarray}
\begin{eqnarray}
B&=&\{ 8,936+\epsilon \delta u(13,899+z^{\text{appl}} \text{cos} \theta (-0,866c_{AB}-0,866c_{AC}-0,732c_{AE}+0,732c_{AD}-0,866c_{AF})) \nonumber\\
&&+\epsilon \delta u z^{\text{appl}} \text{sin} \theta (-0,866d_{AB}+1,5d_{AC}+d_{AE}+d_{AD}+1,5d_{AF})   ) \nonumber\\
&&+z^{\text{appl}} \text{cos} \theta (1,42d_{AB}+2,13d_{AC}+2,13 d_{AD}-1,229c_{AF}) \nonumber\\
&&+z^{\text{appl}} \text{sin} \theta (0,866c_{AB}-1,229c_{AC}+2,258 c_{AD}-2,258c_{AE}-1,229c_{AF}) \nonumber\\
&&+((z^{\text{appl}})^2 \text{cos} \theta \text{sin} \theta) (-c_{AB}d_{AB}-c_{AC}d_{AC}-c_{AD}d_{AD}-c_{AE}d_{AE}-c_{AF}d_{AF})
\end{eqnarray}
and
\begin{eqnarray}
C&=& \{ 13,965+\epsilon \delta u(10,867+z^{\text{appl}} \text{sin} \theta (1,732d_{AB}+1,732_{AC}+1,464d_{AD}+1,464d_{AD}+1,732d_{AF})) \nonumber\\
&&+2,568 (\epsilon \delta u)^2  + z^{\text{appl}} \text{sin} (2,458 d_{AB}+4,516c_{AD}+4,516c_{AE}+2,458d_{AF}) \nonumber\\
&&+(z^{\text{appl}} \text{sin} \theta)^2 (d^2_{AB}+d^2_{AC}+d^2_{AD}+d^2_{AE}+d^2_{AE}) \}. 
\end{eqnarray}
Axial stress component $\sigma_{11}$ has term $(z^{\text{appl}} \text{cos} \theta )^2$ as its higher order term, while shear stress component $\sigma_{12}$ has $(z^{\text{appl}})^2 \text{cos} \theta \text{sin} \theta$ and axial stress component $\sigma_{22}$ has term $(z^{\text{appl}} \text{sin} \theta)^2$. As $\theta$ approaches value of 45$^0$ the shear component becomes greater and at $\theta=45^0$ it takes its maximum value. At this value and also at neighboorhing values, shear stress component $\sigma_{12}$ is of the same order as that of its axial counterparts, $\sigma_{11}, \sigma_{22}$ and cannot be neglected even in a small tension analysis. 

For the $E_{2g}$ mode we have 
\begin{eqnarray}
{\bf z}^*_{AB}&=&(1,42+2\epsilon \delta u-c_{AB} z^{\text{appl}} \text{cos} \theta ) {\bf z}_1+(d_{AB} z^{\text{appl}} \text{sin} \theta) {\bf z}_1, \\
{\bf z}^*_{AC}&=&(2,13+2\epsilon \delta u-c_{AC} z^{\text{appl}} \text{cos} \theta  + 1,5 \epsilon \delta u) {\bf z}_1+(1,229 u+d_{AC} z^{\text{appl}} \text{sin} \theta) {\bf z}_2, \\
{\bf z}^*_{AD}&=&(2,13+ 2\epsilon \delta u+c_{AD} z^{\text{appl}} \text{cos} \theta) {\bf z}_1+(2,258+d_{AD} z^{\text{appl}} \text{sin} \theta) {\bf z}_2, \\
{\bf z}^*_{AE}&=&(2 \epsilon \delta u-c_{AE} z^{\text{appl}} \text{cos} \theta) {\bf z}_1+(2,258+d_{AE} z^{\text{appl}} \text{sin} \theta) {\bf z}_2, \\
{\bf z}^*_{AF}&=&(-1,229+2\epsilon \delta u -c_{AF} z^{\text{appl}} \text{cos} \theta) {\bf z}_1+(1,229+d_{AF} z^{\text{appl}} \text{sin} \theta) {\bf z}_2. 
\end{eqnarray}
The tensor product then reads
\begin{equation}
{\bf z}^* \otimes {\bf z}^*= A {\bf z}_1 \otimes {\bf z}_1 +B({\bf z}_1 \otimes {\bf z}_2+{\bf z}_2 \otimes {\bf z}_1) +C {\bf z}_2 \otimes {\bf z}_2, 
\end{equation}
where
\begin{eqnarray}
A&=& \{ 12,599+\epsilon \delta u(27,636-4z^{\text{appl}} \text{cos} \theta (c_{AB}+c_{AC}+c_{AE}+c_{AD}+c_{AF})) \nonumber\\
&&+17 (\epsilon \delta u)^2  + z^{\text{appl}} \text{cos} (-2,84 c_{AB}-4,26c_{AC}+4,26c_{AD}+2,458c_{AF}) \nonumber\\
&&+(z^{\text{appl}} \text{cos} \theta)^2 (c^2_{AB}+c^2_{AC}+c^2_{AD}+c^2_{AE}+c^2_{AE}) \}, 
\end{eqnarray}
\begin{eqnarray}
B&=&\{ 8,936+\epsilon \delta u(13,948+2z^{\text{appl}} \text{cos} \theta (d_{AB}+d_{AC}+d_{AE}+d_{AD}+d_{AF})) \nonumber\\
&&+z^{\text{appl}} \text{cos} (1,42d_{AB}+2,458d_{AC}+2,13 d_{AD}-1,229d_{AF}) \nonumber\\
&&+z^{\text{appl}} \text{sin} (-1,229c_{AC}+2,258 c_{AD}-2,258c_{AE}-1,229c_{AF}) \nonumber\\
&&+(z^{\text{appl}})^2 \text{cos} \theta \text{sin} \theta (-c_{AB}d_{AB}-c_{AC}d_{AC}-c_{AD}d_{AD}-c_{AE}d_{AE}-c_{AF}d_{AF})
\end{eqnarray}
and
\begin{eqnarray}
C&=& \{ 13,216 + z^{\text{appl}} \text{sin} (2,458 d_{AC}+4,516d_{AD}+4,516c_{AE}+2,458d_{AF}) \nonumber\\
&&+(z^{\text{appl}} \text{sin} \theta)^2 (d^2_{AB}+d^2_{AC}+d^2_{AD}+d^2_{AE}+d^2_{AE}) \}. 
\end{eqnarray}
As for the $A_{1g}$ case, the axial stress component $\sigma_{11}$ has term $(z^{\text{appl}} \text{cos} \theta )^2$ as its higher order term, while shear stress component $\sigma_{12}$ has $(z^{\text{appl}})^2 \text{cos} \theta \text{sin} \theta$ and axial stress component $\sigma_{22}$ has term $(z^{\text{appl}} \text{sin} \theta)^2$. As $\theta$ approaches value of 45$^0$ the shear component becomes greater and at $\theta=45^0$ it takes its maximum value. At this value and also at neighboorhing values, shear stress component $\sigma_{12}$ is of the same order as that of its axial counterparts, $\sigma_{11}, \sigma_{22}$ and cannot be neglected even in a small tension analysis. All in all, the analysis of this Section reveals that shear stress components can be important when an external tensile field applies. 

{\it{Remark}}

For a compressive field a similar analysis can be carried out by altering the sign of quantities accordingly. Nevertheless, care should be taken due to the fact that energetically for a free-standing graphene, buckling modes are preferable and a compressive type of loading produces them at low strain levels.  

\section{Conclusions}

We focus on only one graphene ring and examine to which stress tensor components $E_{2g}$ and $A_{1g}$ modes correspond. For the case when the graphene ring is at rest (i.e. no applied loading), we infer that axial as well as shear stress components are produced as an outcome of the activation of these modes. When a tensile field applies along the armchair direction it is the axial $\sigma_{11}$ Cauchy's stress tensor component which is the dominant one, while when tension is along the zig-zag direction it is $\sigma_{22}$ which dominates. A first approximation analysis is allowable to consider only these quantities, but a full analysis should also take account of the shear $\sigma_{12}$ component which is of smaller order but not negligible. For applied tension along an arbitrary direction describes by angle $\theta$, as $\theta$ approaches value $45^0$, the shear component becomes greater reaching its maximum value at $\theta=45^0$. At this case all stress components, axial $\sigma_{11}$, $\sigma_{22}$ as well as shear $\sigma_{12}$, are of the same order. 

To compare the present framework with the discrete (Newtonian) approach one should introduce plane progressive waves through term $\delta u$ of the perturbation describing the $A_{ag}$ and $E_{2g}$ modes. Such an assumption substituted to the fully dynamical continuum equations would enable to measure the effect of applied stress/strain upon frequency for the continuum approach in parallel with the discrete approach of \cite{Sahataetal1988}. Nevertheless, such a path, from the mathematical point of view is highly non-trivial since one will result to a system of integro-differential equations. Perhaps the recent work of Dayal (\cite{Dayal2017}) studying plane waves for the peridynamic theory, might help as a guide here. When such an analysis is carried out it should match to experimental findings which tell us that the average frequency shifts over an applied (arbitrary) stress/strain are constant regardless of the angle of application $\theta$, for small stress/strain regimes. We conjecture that for larger applied stress/strains it is perhaps reasonable to expect that results will change with $\theta$. 

\section{Acknowledgements}
  
The authors acknowledge the financial support of the European Research Council (ERC Advanced Grant 2013) via project no. 321124, “Tailor Graphene”.  One of us (CG) wishes also to acknowledge the financial support of Graphene FET Flagship (‘‘Graphene- Based Revolutions in ICT And Beyond’’- Grant agreement no: 604391).



\vspace{0.1cm}

D. Sfyris\\
National Technical University of Athens\\
Scholl of Applied Mathematics and Physics\\
Section of Mechanics\\
Athens, Greece \\
and FORTH/ICE-HT, Patras, Greece \\
dsfyris@sfyris.net\\
www.sfyris.net

\vspace{0.2cm}
G.I. Sfyris\\
University of Piraeus\\
Piraeus, Greece \\
gsfyris@gmail.com \\
www.sfyris.net

\vspace{0.2cm}
C. Galiotis \\
FORTH/ICE-HT, Patras, Greece \\
and Department of Chemical Engineering \\
University of Patras \\
Patras, Greece \\
c.galiotis@iceht.forth.gr \\
galiotis@chemeng@upatras.gr

\end{document}